\pdfoutput=1
\documentclass[aps, twocolumn, floatfix, superscriptaddress]{revtex4-1}
\usepackage{graphicx}
\DeclareGraphicsExtensions{.pdf}
\usepackage{amsmath,amssymb,bbold,bm,color}
\usepackage{float}
\usepackage{epstopdf}
\usepackage{tabularx}
\usepackage{braket}
\usepackage{booktabs}
\usepackage{array}
\usepackage{blindtext}
\usepackage{gensymb}
\usepackage{dutchcal}
\usepackage[colorlinks=true,breaklinks=true,linkcolor=blue,anchorcolor=red,citecolor=blue,urlcolor=blue]{hyperref}
\usepackage{seqsplit}
\allowdisplaybreaks[3]

\begin{document}

\title{Dispersions and magnetism of strain-induced pseudo Landau levels in Bernal-stacked bilayer graphene}

\author{Tianyu Liu}
\affiliation{International Quantum Academy, Shenzhen 518048, China}
\affiliation{Shenzhen Key Laboratory of Quantum Science and Engineering, Shenzhen 518055, China}
\affiliation{Max Planck Institute for the Physics of Complex Systems, 01187 Dresden, Germany}

\author{Jun-Hong Li}
\affiliation{Department of Physics and Guangdong Basic Research Center of Excellence for Quantum Science, Southern University of Science and Technology (SUSTech), Shenzhen 518055, China}
\affiliation{Quantum Science Center of Guangdong-Hong Kong-Macao Greater Bay Area (Guangdong), Shenzhen 518045, China}

\author{Xingchuan Zhu}
\affiliation{School of Intelligent Manufacturing, Nanjing University of Science and Technology, Nanjing 210094, China }


\author{Huaiming Guo}
\affiliation{ School of Physics, Beihang University,
Beijing 100191, China}

\author{Hai-Zhou Lu}
\email{luhz@sustech.edu.cn}
\affiliation{Department of Physics and Guangdong Basic Research Center of Excellence for Quantum Science, Southern University of Science and Technology (SUSTech), Shenzhen 518055, China}
\affiliation{Quantum Science Center of Guangdong-Hong Kong-Macao Greater Bay Area (Guangdong), Shenzhen 518045, China}

\author{X. C. Xie}
\affiliation{Interdisciplinary Center for Theoretical Physics and Information
Sciences (ICTPIS), Fudan University, Shanghai 200433, China}
\affiliation{International Center for Quantum Materials, School of Physics, Peking University, Beijing 100871, China}
\affiliation{Hefei National Laboratory, Hefei 230088, China}

\begin{abstract} 
Elastic strain can displace the massless Dirac fermions in monolayer graphene in a space-dependent fashion, similar to the effect of an external magnetic field, thus giving rise to Landau quantization. We here show that the strain-induced Landau quantization can also take place in Bernal-stacked bilayer graphene, where the low-energy excitations are massive rather than Dirac-like. The zigzag ribbon of Bernal-stacked bilayer graphene realizes a two-legged Su-Schrieffer-Heeger model with a domain wall, which coincides with the guiding center of the strain-induced pseudo Landau levels. We reduce the lattice model of the ribbon in the vicinity of the guiding center into an exactly solvable coupled Dirac model and analytically derive the dispersions of the strain-induced pseudo Landau levels. Remarkably, the zeroth and first pseudo Landau levels are dispersionless and sublattice-polarized. We elucidate that the interaction on these two pseudo Landau levels results in a global antiferromagnetic order. Our study extends the strain-induced Landau quantization to the massive excitations and indicates strain as a tuning knob of magnetism.
\end{abstract}

\date{\today}
\maketitle

\section{Introduction}
One of the most intriguing properties of monolayer graphene is that its Landau quantization does not necessarily require magnetic fields as long as proper nonuniform elastic strain is applied \cite{guinea2010a, levy2010, vozmediano2010}. This is because the massless Dirac fermions of monolayer graphene are displaced in the momentum space by the strain in a space-dependent fashion, similar to the magnetic field effect (i.e., the Peierls substitution). In fact, the strain-induced Landau quantization is only associated with the Diracness of the low-energy excitations, regardless of their types. The resulting strain-induced pseudo Landau levels have been found in Dirac/Weyl semimetals \cite{cortijo2015, cortijo2016, sumiyoshi2016, pikulin2016, grushin2016, arjona2017, liu2017a, ilan2020, lee2022}, Dirac/Weyl superconductors \cite{liu2017b, matsushita2018}, $d$-wave superconductors \cite{massarelli2017, nica2018}, kagome crystals \cite{liu2020}, Kitaev quantum spin liquids \cite{rachel2016, perreault2017}, honeycomb magnets \cite{ferreiros2018, liu2019, nayga2019, liu2021, sunjunsong2021a, sunjunsong2021b, liu2023}, graphene-like and Weyl photonic devices \cite{rechtsman2013, jia2019, bellec2020, jamadi2020}, synthetic $\alpha-T_3$ lattice and brick-wall lattice materials \cite{sunjunsong2022, filusch2022, sunjunsong2023}, and acoustic Weyl metamaterials \cite{yang2017, abbaszadeh2017, brendel2017, wen2019, peri2019, cheng2024}.

For monolayer graphene, the pseudo Landau levels induced by experimentally accessible strain patterns are in general dispersive except for the zeroth one \cite{shi2021, liu2022}, which is topologically protected to be dispersionless and thus can produce interaction effect \cite{roy2014, elias2023}. The interaction leads to an edge-compensated global antiferromagnetic order, which comprises opposite ferromagnetic orders carried by the sublattice-polarized zeroth pseudo Landau level \cite{shi2021, liu2022, settnes2016} in the bulk and the topological edge state terminated on a different sublattice \cite{roy2014, elias2023}. 

Bilayer graphene with Bernal stacking sequence \cite{mccann2006, ohta2006}, i.e., half of the carbon atoms of one monolayer lying over the centers of the hexagonal plaquettes of the other monolayer [Fig.~\ref{fig1}(a)], hosts massive fermions in contrast to the massless Dirac fermions in monolayer graphene \cite{mccann2006, ohta2006, novoselov2006, geim2007}. It is recently realized that strain can possibly couple to such massive excitations as a gauge field  \cite{wan2023}. However, detailed characterization of the induced pseudo Landau levels is lacking and the interaction-induced magnetism carried by such pseudo Landau levels is also elusive.

In this paper, we analytically derive the dispersions of the strain-induced pseudo Landau levels in bent and twisted ribbons of Bernal-stacked bilayer graphene and show that interaction can produce global antiferromagnetism with edge compensation. In Sec.~\ref{sec2}, we briefly review the bulk and ribbon models of the strain-free Bernal-stacked bilayer graphene, paying close attention to the electronic structures. In Sec.~\ref{sec3}, we elucidate that the bent/twisted Bernal-stacked bilayer graphene ribbon at a given momentum can be understood as a two-legged Su-Schrieffer-Heeger model exhibiting at the charge neutrality point topological end and domain-wall modes. We further demonstrate that the domain-wall modes turn out to be the zeroth and first pseudo Landau levels and derive the dispersions of all the pseudo Landau levels by reducing the ribbon model in the vicinity of the domain wall (i.e., the guiding center of the pseudo Landau levels) into an analytically solvable coupled Dirac model. We find that the dispersionless zeroth (first) pseudo Landau level only resides on one (two) of the four sublattices, exhibiting sublattice polarization. In Sec.~\ref{sec4}, we model the interaction with a Hubbard model. At the mean-field level, we find opposite ferromagnetic orders carried by the sublattice-polarized zeroth and first pseudo Landau levels and the topological edge states terminated on the other sublattices. Section~\ref{sec5} concludes the paper and outlooks a few promising future directions.

\section{Model}
\label{sec2}
\begin{figure*}[t]
\includegraphics[width = \textwidth]{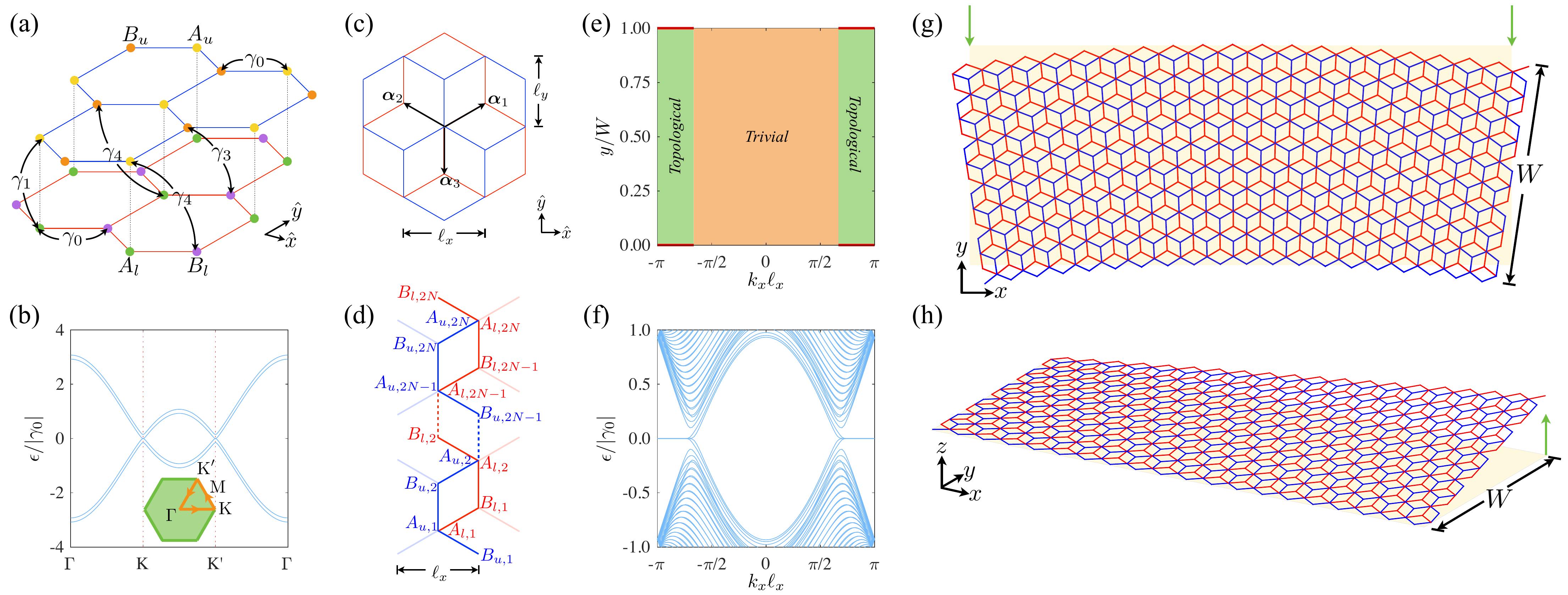}
\caption{(a) Schematic plot of the lattice of the strain-free Bernal-stacked bilayer graphene with various hoppings labeled. (b) Bulk bands of the strain-free Bernal-stacked bilayer graphene plotted along the high-symmetry path. Inset: Brillouin zone with high-symmetry points and path (orange) labeled. (c) Top view of the strain-free Bernal-stacked bilayer graphene with $(\bm \alpha_1, \bm \alpha_2, \bm \alpha_3)$ being the intralayer nearest-neighbor vectors. (d) Unit cell of the zigzag ribbon of the strain-free Bernal-stacked bilayer graphene. The solid (partially transparent) lines correspond to intra-unit-cell (inter-unit-cell) chemical bonds. The bonds associated with the lower (upper) honeycomb monolayer are colored red (blue). (e) Phase diagram of the zigzag ribbon of the strain-free Bernal-stacked bilayer graphene. The ribbon Hamiltonian realizes a topological Su-Schrieffer-Heeger model for $\tfrac{2\pi}{3\ell_x} \leq |k_x| \leq \tfrac{\pi}{\ell_x}$ (green) and a trivial Su-Schrieffer-Heeger model for $|k_x|\leq \tfrac{2\pi}{3\ell_x}$ (orange). The red bold lines at $y=0$ and $y=W$ highlight the regimes of the lower (on sites $A_{l,1}$ and $B_{u,1}$) and upper (on sites $B_{l,2N}$ and $A_{u,2N}$) edge states. (f) Spectrum of the zigzag ribbon of the strain-free Bernal-stacked bilayer graphene. The four-fold zero-energy flat band in the regime $\tfrac{2\pi}{3\ell_x}\leq|k_x|\leq\tfrac{\pi}{\ell_x}$ represents the edge states localized at sites $A_{l,1}$, $B_{l,2N}$, $B_{u,1}$, and $A_{u,2N}$. (g) Circularly bent ribbon of Bernal-stacked bilayer graphene produced by applying in-plane transverse forces (marked by the green arrows). (h) Twisted ribbon of Bernal-stacked bilayer graphene formed by lifting up a corner with an out-of-plane force (marked by the green arrow). In panels (g) and (h), the strained lower (upper) honeycomb monolayer is colored red (blue). For clarity, the undeformed ribbons are represented by the yellow shades and their lattices are not explicitly illustrated. In panels (b) and (f), $\gamma_1=0.15|\gamma_0|$ is used. This parameter setting is consistent with various numerical and experimental works \cite{partoens2006, min2007, malard2007, zhang2008, gruneis2008, nilsson2008, gava2009, kuzmenko2009, jung2014}.
} \label{fig1}
\end{figure*}

We start by briefly reviewing the tight-binding model of the strain-free Bernal-stacked bilayer graphene, comprising of two honeycomb monolayers [Fig.~\ref{fig1}(a)] of the same lattice constant $a=1.42\,\mathring{\text{A}}$ \cite{saito1998, dresselhaus2002}. The sublattices in the lower (upper) honeycomb monolayer are labeled as $A_l$ ($A_u$) and $B_l$ ($B_u$), respectively. In the $x$-$y$ plane, the $A_l$ sites coincide with the $A_u$ sites, while the $B_l$ ($B_u$) sites sit at the centers of the hexagonal plaquettes of the upper (lower) honeycomb monolayer [Figs.~\ref{fig1}(a, c)]. For this reason, $A_l$ and $A_u$ sites are referred to as the dimer sites, while $B_l$ and $B_u$ sites are called non-dimer sites \cite{mccann2013}. The nearest-neighbor vectors are $(\bm \alpha_1, \bm \alpha_2, \bm \alpha_3)=(\tfrac{1}{2}\ell_x \hat x+\tfrac{1}{3}\ell_x\hat y, -\tfrac{1}{2}\ell_x\hat x+\tfrac{1}{3}\ell_y\hat y, -\tfrac{2}{3}\ell_y\hat y)$, where we set $\ell_x=\sqrt 3 a$ and $\ell_y=\tfrac 3 2 a$ [Fig.~\ref{fig1}(c)].  

In this section, we study the strain-free Bernal-stacked bilayer graphene with tight-binding models enclosing only the nearest-neighbor intralayer and interlayer hoppings, while a more realistic model with more hoppings considered is given in Appendix~\ref{app_a}.

\subsection{Periodic model}
\label{sec2a}
For the periodic (i.e., infinite) Bernal-stacked bilayer graphene, we adopt the spinless tight-binding Hamiltonian 
\begin{equation} \label{H1}
\begin{split}
H_1=\sum_{\bm r,j} \gamma_0 ( a_{u, \bm r}^\dagger b_{u, \bm r-\bm\alpha_j} &+ a_{l, \bm r}^\dagger b_{l, \bm r+\bm \alpha_j}) 
\\
+ \sum_{\bm r}\gamma_1 a_{l, \bm r}^\dagger a_{u, \bm r} &+ \text{H.c.},
\end{split}
\end{equation}
where $a_{l(u),\bm r}$ and $b_{l(u),\bm r}$ are the annihilation operators of the two sublattices in the lower (upper) honeycomb monolayer; and $\gamma_0$ and $\gamma_1$ are respectively the intralayer and interlayer nearest-neighbor hoppings [Fig.~\ref{fig1}(a)]. 

We perform a Fourier transform $[a_{l(u), \bm r}, b_{l(u), \bm r}]^T=N^{-1/2}\sum_{\bm k} e^{i\bm k\cdot\bm r}[a_{l(u), \bm k}, b_{l(u), \bm k}]^T$, where $N$ is the number of unit cells. The nearest-neighbor tight-binding Hamiltonian can then be rewritten in a matrix product form as $H_1=\sum_{\bm k} \Psi_{\bm k}^\dagger \mathcal H_{1,\bm k} \Psi_{\bm k}$, where the basis reads $\Psi_{\bm k}=(a_{l,\bm k}, b_{l,\bm k}, a_{u,\bm k}, b_{u,\bm k})^T$ and the first quantized Bloch Hamiltonian reads
\begin{equation} \label{H1k}
\mathcal H_{1,\bm k}=\gamma_0\Re(f_{\bm k})\sigma^x-\gamma_0\Im(f_{\bm k}) \sigma^y\tau^z+\frac{1}{2}\gamma_1\tau^x + \frac{1}{2}\gamma_1\sigma^z\tau^x,
\end{equation}
%
where we have defined a parameter $f_{\bm k} = \sum_j e^{i \bm k \cdot \bm \alpha_j}$; and $\bm \sigma$ and $\bm \tau$ are Pauli matrices respectively defined in the sublattice and layer spaces. The spectrum of $\mathcal H_{1,\bm k}$ can be easily solved  as
\begin{equation} \label{E1}
\mathcal E_{\bm k}= \eta \sqrt{\gamma_0^2|f_{\bm k}|^2+\frac{\gamma_1^2}{4}} +\xi \frac{\gamma_1}{2},
\end{equation}
where $\eta=\pm 1$ and $\xi=\pm 1$ give in total four energy bands. In the vicinity of the Brillouin zone corners $\bm k_s=(s \tfrac{4\pi}{3\sqrt 3 a}, 0)$  [Inset, Fig.~\ref{fig1}(b)], where $s=\pm 1$, we have $f_{\bm k_s +\bm q}= -\tfrac{3}{2}a(s q_x + i q_y)$. Thus, the dispersion [Eq.~(\ref{E1})] is reduced to $\epsilon_{\bm k_s+\bm q} = (\eta+\xi) \tfrac{\gamma_1}{2}+ \eta \hbar^2 v_F^2 q^2/\gamma_1$, where $v_F= \tfrac{3a}{2\hbar}|\gamma_0|$. This means that all the energy bands are quadratically dispersing with an effective mass $m^*=\eta \gamma_1/(2v_F^2)$. The outer two bands are separated by a gap of $2\gamma_1$, while the inner two bands touch at the Brillouin zone corners [Fig.~\ref{fig1}(b)].

\subsection{Ribbon model}
\label{sec2b}
We now consider a zigzag ribbon of the strain-free Bernal-stacked bilayer graphene along the $x$ direction. We perform to the two-dimensional tight-binding Hamiltonian [Eq.~(\ref{H1})] a partial Fourier transform $[a_{l(u), \bm r}, b_{l(u), \bm r}]^T= \tilde N^{-1/2}\sum_{k_x} e^{i k_xx}  [a_{l(u), k_x, y}, b_{l(u), k_x, y}]^T$, where $\tilde N$ is the number of ribbon unit cells [Fig.~\ref{fig1}(d)], and obtain a ribbon tight-binding Hamiltonian
\begin{align} \label{H1rib}
H_1= &\sum_{k_x,y} a_{u,k_x,y}^\dagger  \left[2\gamma_0 \cos\left(\frac{1}{2} k_x \ell_x \right) + \gamma_0 \hat s_{\ell_y}\right] b_{u,k_x,y-\frac{\ell_y}{3}} \nonumber
\\
+&\sum_{k_x,y} a_{l,k_x,y}^\dagger  \left[2\gamma_0 \cos\left(\frac{1}{2} k_x \ell_x \right) + \gamma_0 \hat s_{-\ell_y}\right] b_{l,k_x,y+\frac{\ell_y}{3}} \nonumber
 \\
+&\sum_{k_x,y} \gamma_1 a_{l,k_x,y}^\dagger a_{u,k_x,y} + \text{H.c.},
\end{align}
where $\hat s_{\pm \ell_y}$ is the $y$-direction shift operator satisfying $\hat s_{\pm \ell_y} b_{k_x,y}=b_{k_x,y \pm \ell_y}$. Equation~(\ref{H1rib}) may also be constructed by directly writing down the tight-binding Hamiltonian of the ribbon unit cell [Fig.~\ref{fig1}(d)]. 

When the two honeycomb monolayers are decoupled (i.e., $\gamma_1=0$), Eq.~(\ref{H1rib}) at a given momentum $k_x$ characterizes two independent Su-Schrieffer-Heeger (SSH) chains \cite{su1979}. Each of the chains has an \emph{intracell} hopping $2\gamma_0\cos(\tfrac{1}{2}k_x\ell_x)$ and an \emph{intercell} hopping $\gamma_0$. The topology of the SSH chains thus relies on the momentum $k_x$. Specifically, for $\tfrac{2\pi}{3\ell_x} \leq |k_x| \leq \tfrac{\pi}{\ell_x}$ ($0\leq |k_x| \leq \tfrac{2\pi}{3\ell_x}$), the intercell (intracell) hopping dominates and both chains are topological (trivial) [Fig.~\ref{fig1}(e)]. For a fixed $k_x$ in the topological regime, the two SSH chains contribute in total four zero-energy end modes, which are protected by chiral symmetry $\mathcal S=\tau^z\sigma^z$ and are respectively located on sites $A_{l,1}$, $B_{l,2N}$, $B_{u,1}$, and $A_{u,2N}$. The two end modes located at sites $A_{l,1}$ and $B_{u,1}$ ($B_{l,2N}$ and $A_{u,2N}$) at different $k_x$'s constitute two flat bands at the charge neutrality point, corresponding to the lower (upper) edge states. There are thus in total four zero-energy flat bands associated with the four edges (i.e., two for each monolayer) of the strain-free Bernal-stacked bilayer graphene.

For $\gamma_1\neq 0$, Eq.~(\ref{H1rib}) characterizes two SSH chains coupled by the interlayer nearest-neighbor hopping, which preserves the chiral symmetry. The resulting two-legged SSH chain thus can be topologically connected to the aforementioned decoupled model. And we should expect an intact phase diagram [Fig.~\ref{fig1}(e)] and four similar zero-energy edge bands [Fig.~\ref{fig1}(f)].

\section{Strain-induced pseudo Landau levels}
\label{sec3}
We here consider two strain patterns exerted to a zigzag ribbon of Bernal-stacked bilayer graphene: (i) An in-plane circular bend that may be achieved by applying transverse forces to the ribbon [Fig.~\ref{fig1}(g)]; and (ii) An out-of-plane twist that may be achieved by lifting up a single corner of the ribbon, while leaving the other three corners co-planar [Fig.~\ref{fig1}(h)]. These two strain patterns only break the translational symmetry in the $y$ direction, thus are compatible with the ribbon geometry. We here adopt the nearest-neighbor ribbon model discussed in Sec.~\ref{sec2b}.  The strain effects beyond the nearest-neighbor model will be discussed in Appendix~\ref{app_b}. 

\subsection{Models of strained ribbons}
\label{sec3a}
For simplicity, we here assume the both strain patterns do not alter the thickness of the bilayer graphene so that $\gamma_1$ is kept as a constant. On the other hand, the intralayer spacing between any two nearest neighbors is spatially modulated by strain according to
\begin{equation} \label{sub}
\gamma_0(\bm r,\bm r')=\gamma_0 \exp \left( -g\frac{|\bm r'-\bm r|-a}{a}\right),
\end{equation}
where $\bm r$ and $\bm r'$ mark the positions of the nearest neighbors; $g$ is the Gr\"uneisen parameter; and $\gamma_0$ is the strain-free nearest-neighbor hopping. The exponential form [Eq.~(\ref{sub})] adopted here has been widely used in the straintronics of monolayer graphene \cite{pereira2009}. In Eq.~(\ref{sub}), the strain-modulated nearest-neighbor spacing $|\bm r'-\bm r|$ is found to be $[\alpha_{i,x}^2(1+\lambda_by)^2+\alpha_{i,y}^2]^{1/2}$ \cite{liu2022} for circular bend [Fig.~\ref{fig1}(g)] and $[\alpha_{i,x}^2(1+\lambda_t^2y^2)+\alpha_{i,y}^2]^{1/2}$ \cite{shi2021} for twist [Fig.~\ref{fig1}(h)], where $\lambda_b>0$ and $\lambda_t >0$ respectively scales the tensile strain in bent and twisted ribbons. Therefore, the hoppings along bonds $\pm\bm\alpha_3$ are still $\gamma_0$ for both strain patterns, while the hoppings along $\pm\bm\alpha_{1,2}$ for circular bend and twist are respectively 
\begin{equation} \label{gamma0bt}
\begin{split}
\gamma_0^b(y) &=\gamma_0 \exp \left[ g\left( 1-\sqrt{1+\frac 3 4 \lambda_b^2 y^2 + \frac{3}{2}\lambda_b y} \right) \right], 
\\
\gamma_0^t(y) &=\gamma_0 \exp \left[ g\left( 1-\sqrt{1+\frac 3 4 \lambda_t^2 y^2} \right) \right], 
\end{split}
\end{equation}
where the superscripts $b$ and $t$ respectively stand for bend and twist. Upon on plugging into the tight-binding model [Eq.~(\ref{H1})] and performing a partial Fourier transform in the $x$ direction, we obtain a ribbon tight-binding Hamiltonian  
\begin{align} \label{H1bt}
H_1^{b,t}= &\sum_{k_x,y} a_{u,k_x,y}^\dagger  \left[2\gamma_0^{b,t}(y) \cos\left(\frac{1}{2} k_x \ell_x \right) + \gamma_0 \hat s_{\ell_y}\right] b_{u,k_x,y-\frac{\ell_y}{3}} \nonumber
\\
+&\sum_{k_x,y} a_{l,k_x,y}^\dagger  \left[2\gamma_0^{b,t}(y) \cos\left(\frac{1}{2} k_x \ell_x \right) + \gamma_0 \hat s_{-\ell_y}\right] b_{l,k_x,y+\frac{\ell_y}{3}} \nonumber
 \\
+&\sum_{k_x,y} \gamma_1 a_{l,k_x,y}^\dagger a_{u,k_x,y} + \text{H.c.},
\end{align}
which, similar to Eq.~(\ref{H1rib}), characterizes two coupled SSH chains for a given momentum $k_x$. The only difference is that the intracell hopping $2\gamma_0^{b,t}(y)\cos(\tfrac{1}{2}k_x\ell_x)$ is now space-dependent because of the strain modulation. Consequently, domain walls emerge at
\begin{align} \label{domain}
l_0^b &=\frac{1}{\lambda_b} \sqrt{\frac{4}{3}\left\{1+\frac{1}{g }\ln \left[ 2\cos \left( \frac{1}{2}k_x\ell_x \right) \right] \right\}^2 -\frac{1}{3}} -\frac{1}{\lambda_b} , \nonumber
\\
l_0^t &= \frac{1}{\lambda_t} \sqrt{\frac{4}{3}\left\{1+\frac{1}{g }\ln \left[ 2\cos \left( \frac{1}{2}k_x\ell_x \right) \right] \right\}^2 -\frac{4}{3}},
\end{align}
for the bent [Fig.~\ref{fig1}(g)] and twisted [Fig.~\ref{fig1}(h)] ribbons, respectively.

\subsection{Phase diagrams of strained ribbons}
\label{sec3b}
The appearance of the domain walls [Eq.~(\ref{domain})] implies that the phase diagrams of the bent and twisted ribbons  are more complicated than their strain-free counterparts [Fig.~\ref{fig1}(e)]. We emphasize that these phase diagrams are sensitive to the values of $\lambda_{b,t}$. Setting $2\gamma_0^{b,t}(W)=\gamma_0$, where $W$ is the ribbon width, we find the following critical values 
\begin{equation}
\begin{split}
\lambda_b^{\text c} &=\frac{1}{W} \left[ \sqrt{\frac{4}{3} \left(1+\frac{1}{g}\ln 2 \right)^2-\frac{1}{3}}-1\right],
\\
\lambda_t^{\text c} &=\frac{1}{W} \sqrt{\frac{4}{3} \left(1+\frac{1}{g}\ln 2 \right)^2-\frac{4}{3}},
\end{split}
\end{equation}
with which the intercell hopping $\gamma_0$ always dominates the intracell hopping $2\gamma_0^{b,t}(W)\cos(\tfrac 1 2 k_x\delta_x)$ at the upper (i.e., $y=W$) edge.

\begin{figure*}[t]
\includegraphics[width = \textwidth]{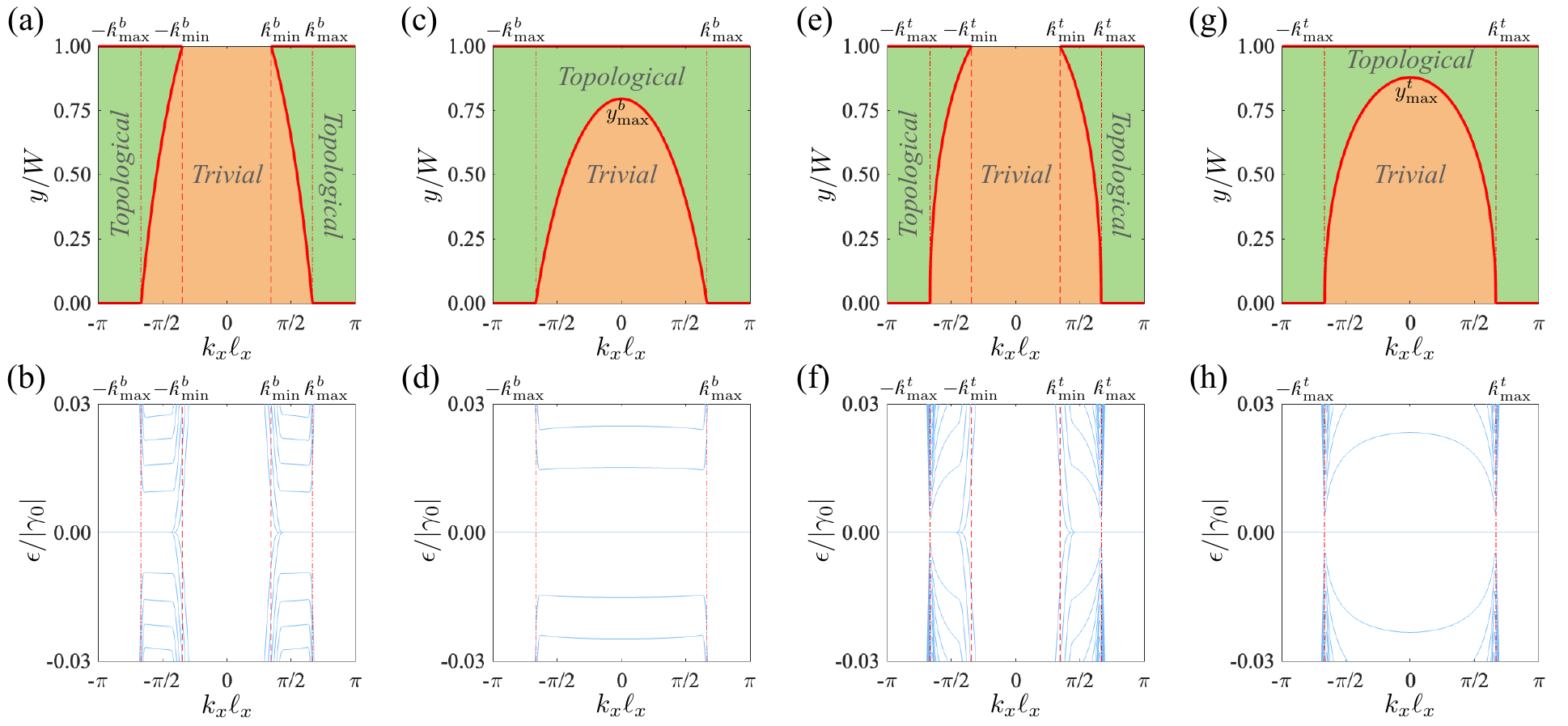}
\caption{ Phase diagrams and low-energy spectra of strained ribbons of Bernal-stacked bilayer graphene. In panels (a-d), dot-dashed and dashed lines respectively mark $\pm \mathcal k_{\max}^b$ and $\pm \mathcal k_{\min}^b$ with $\mathcal k_{\max}^b=\tfrac{2\pi}{3\ell_x}$ and $\mathcal k_{\min}^b= \tfrac{2}{\ell_x}\arccos\{ \tfrac{1}{2}\exp[ g(1+\tfrac{3}{2}\lambda_bW + \tfrac{3}{4}\lambda_b^2W^2)^{1/2}-g ] \}$. In panels (e-h), the dot-dashed and dashed lines respectively mark $\pm \mathcal k_{\max}^t$ and $\pm \mathcal k_{\min}^t$ with $\mathcal k_{\max}^t=\tfrac{2\pi}{3\ell_x}$ and $\mathcal k_{\min}^t=\tfrac{2}{\ell_x}\arccos\{\tfrac{1}{2}\exp[g(1+\tfrac{3}{4} \lambda_t^2W^2) ^{1/2}-g]\}$. The critical values of $\lambda$ for bent and twisted ribbons are $\lambda_b^{\text c}=0.267W^{-1}$ and  $\lambda_t^{\text c}=0.777W^{-1}$, respectively. (a) Phase diagram for bend $\lambda_b=0.209W^{-1}<\lambda_b^{\text c}$. The longer (shorter) red line segments highlight the momentum regimes of the SSH edge modes located at the top (bottom) edges. The red curves, which are determined by Eq.~(\ref{domain}), mark the SSH domain-wall modes (i.e., the zeroth and first pseudo Landau levels) separating the topological (green) and trivial (orange) sectors. (b) Spectrum for bend $\lambda_b=0.209W^{-1}<\lambda_b^{\text c}$ obtained by diagonalizing Eq.~(\ref{H1bt}). Panels (c, d) are similar to (a, b) but have a stronger bend $\lambda_b=0.336W^{-1}>\lambda_b^{\text c}$. The phase boundary between the topological and trivial sectors is peaked at $y_{\max}^b=0.794W$. Panels (e, f) are similar to (a, b) but characterize a weakly twisted ribbon with $\lambda_t=0.679W^{-1}<\lambda_t^{\text c}$, whose generated edge strain is the same as $\lambda_b=0.209W^{-1}$. Panels (g, h) are similar to (a, b) but characterize a strongly twisted ribbon with $\lambda_t=0.885W^{-1}>\lambda_t^{\text c}$, whose generated edge strain is the same as $\lambda_b=0.336W^{-1}$. The phase boundary between the topological and trivial sectors is peaked at $y_{\max}^t=0.879W$. In all panels, $\gamma_1=0.15|\gamma_0|$ is used.
 } \label{fig2}
\end{figure*}

For weak strain, i.e., $\lambda<\lambda_{b,t}^\text{c}$, we find the domain walls are bounded inside the ribbons for the momenta $\mathcal k_{\min}^{b,t}\leq |k_x| \leq \mathcal k_{\max}^{b,t}$ with $\mathcal k_{\min}^b= \tfrac{2}{\ell_x}\arccos\{ \tfrac{1}{2}\exp[ g(1+\tfrac{3}{2}\lambda_bW + \tfrac{3}{4}\lambda_b^2W^2)^{1/2}-g ] \}$, $\mathcal k_{\min}^t=\tfrac{2}{\ell_x}\arccos\{\tfrac{1}{2}\exp[g(1+\tfrac{3}{4} \lambda_t^2W^2) ^{1/2}-g]\}$, and $\mathcal k_{\max}^b=\mathcal k_{\max}^t =\tfrac{2\pi}{3\ell_x}$. $\mathcal k_{\max}^{b,t}$ and $\mathcal k_{\min}^{b,t}$ are determined by setting $l_0^{b,t}=0$ and $l_0^{b,t}=W$ in Eq.~(\ref{domain}), respectively. For a given momentum inside this regime [between the dot-dashed and dashed lines in Figs.~\ref{fig2}(a, e)], the domain walls [Eq.~(\ref{domain})] divide the coupled SSH chains into a topological upper sector and a trivial lower sector. Consequently, there exist two zero-energy SSH end modes at $y=W$ on sites $B_{l,2N}$ and $A_{u,2N}$ and two zero-energy SSH domain-wall modes at $y=l_0^{b,t}$ associated with sublattices $A_l$ and $B_u$ [Figs.~\ref{fig2}(a, e)]. On the other hand, at a given momentum where the domain walls have no supports, the coupled SSH chains must be entirely topological or trivial. Specifically, for $\mathcal k_{\max}^{b,t}<|k_x|<\tfrac{\pi}{\ell_x}$, the intercell hopping dominates, giving rise to two coupled topological SSH chains with two zero-energy SSH end modes at $y=W$ on sites $B_{l,2N}$ and $A_{u,2N}$ and another two at $y=0$ on sites $A_{l,1}$ and $B_{u,1}$  [Figs.~\ref{fig2}(a, e)]. In the meanwhile, for $|k_x|<\mathcal k_{\min}^{b,t}$, intracell hopping dominates and no SSH end modes exist. 

The aforementioned zero-energy SSH end and domain-wall modes result in at the charge neutrality point several flat bands: (i) two bulk bands at $l_0^{b,t}$ in the regime $\mathcal k_{\min}^{b,t}\leq |k_x| \leq \mathcal k_{\max}^{b,t}$; (ii) two edge bands at $y=0$ in the regime $\mathcal k_{\max}^{b,t}\leq |k_x| \leq \tfrac{\pi}{\ell_x}$; and (iii) two edge bands at $y=W$ in the regime $\mathcal k_{\min}^{b,t}\leq |k_x| \leq \tfrac{\pi}{\ell_x}$. To substantiate this claim, we perform numerical diagonalization of the tight-binding Hamiltonian [Eq.~(\ref{H1bt})]. Indeed, we find four zero-energy flat bands in the expected momentum regimes except that these bands acquire dispersions before reaching $\mathcal k_{\min}^{b,t}$ [Figs.~\ref{fig2}(b, f)]. This is because both the SSH end and domain-wall modes are not perfectly localized and thus they get overlapped before reaching $y=W$, corresponding to a momentum slightly away from $\mathcal k_{\min}^{b,t}$, according to Eq.~(\ref{domain}).

For strong strain, i.e., $\lambda>\lambda_{b,t}^{\text c}$, $\mathcal k_{\min}^{b,t}$ is no longer well defined. The disappearance of the lower bounds suggests that the domain walls stay inside the ribbons for $|k_x|\leq \mathcal k_{\max}^{b,t}$ [between the two dot-dashed lines in Figs.~\ref{fig2}(c, g)]. For a given $k_x$ in this regime, the coupled SSH chains [Eq.~(\ref{H1bt})] are divided into an upper topological sector and a lower trivial sector. Therefore, there are two zero-energy SSH end modes at $y=W$ on sites $B_{l,2N}$ and $A_{u,2N}$ and two zero-energy SSH domain-wall modes at $y=l_0^{b,t}$ associated with sublattices $A_l$ and $B_u$ [Figs.~\ref{fig2}(c, g)]. For a given $k_x$ outside this regime, the coupled SSH chains [Eq.~(\ref{H1bt})] become entirely topological, exhibiting two zero-energy SSH end modes at $y=W$ on sites $B_{l,2N}$ and $A_{u,2N}$ together with two end modes at $y=0$ on sites $A_{l,1}$ and $B_{u,1}$ [Figs.~\ref{fig2}(c, g)].

The above zero-energy SSH end and domain-wall modes produce at the charge neutrality point several flat bands: (i) two bulk bands at $l_0^{b,t}$ in the regime $|k_x| \leq \mathcal k_{\max}^{b,t}$; (ii) two edge bands at $y=0$ in the regime $\mathcal k_{\max}^{b,t}\leq |k_x|\leq \tfrac{\pi}{\ell_x}$; and (iii) two edge bands at $y=W$ traversing the whole Brillouin zone. These claims are verified through numerical diagonalization, where four flat bands are observed traversing the whole Brillouin zone as expected [Figs.~\ref{fig2}(d, h)].

\subsection{Coupled Dirac model}
\label{sec3c}
Among all the flat bands at the charge neutrality point, we are most interested in the bulk ones. These bands arise from the SSH domain-wall modes, and thus are localized (around $l_0^{b,t}$) and topological. In strained Dirac matter, the only states that can simultaneously exhibit bulk nature, localization feature, and topology are the pseudo Landau levels. Considering that the two bulk bands resulting from the SSH domain-wall modes are at the charge neutrality point, we should interpret them as the zeroth and first pseudo Landau levels. Therefore, the domain wall $l_0^{b,t}$ should be understood as the guiding center of the strain-induced pseudo Landau levels. As the pseudo Landau levels are all well localized states, to investigate their dispersions, it would be sufficient to study the strained ribbon model [Eq.~(\ref{H1bt})] in the vicinity of the guiding center. 

We first focus on the kernel of Eq.~(\ref{H1bt}), which, in the basis $\Psi_{k_x,y}=(a_{l,k_x,y}, b_{l,k_x,y}, a_{u,k_x,y},b_{u,k_x,y})^T$, reads
\begin{align} \label{H1kxbt}
\mathcal H_{1,k_x,y}^{b,t} = &\left[2\gamma_0^{b,t}(y)\cos\left(\frac{1}{2}k_x\ell_x\right)+\gamma_0\right]\sigma^x - i\gamma_0\ell_y \frac{d}{dy}\sigma^y\tau^z \nonumber
\\
&+\frac{1}{2}\gamma_1\tau^x+\frac{1}{2}\gamma_1\sigma^z\tau^x.
\end{align}
Note that in Eq.~(\ref{H1kxbt}), we have taken the continuum limit such that $\hat s_{\pm \ell_y}\approx 1\pm \ell_y \tfrac{d}{dy}$, which should be legitimate because elastic strain always varies slowly on the lattice scale. To extract the dispersions of the pseudo Landau levels, we study the spectrum of $\mathcal H_{1,k_x,y}^{b,t}$ around the guiding center $l_0^{b,t}$, while the generic spectrum is hardly solvable in an analytically rigorous fashion because of the complicated $y$ dependence of $\gamma_0^{b,t}(y)$.

Before we linearize $\mathcal H_{1,k_x,y}^{b,t}$ in the vicinity of the guiding center, we first determine the momentum regime in consideration. For convenience, we choose to work in the regime $\tfrac{\pi}{\ell_x}\leq k_x \leq \tfrac{3\pi}{\ell_x}$ and then map back to the conventional first Brillouin zone, i.e., $-\tfrac{\pi}{\ell_x} \leq k_x \leq \tfrac{\pi}{\ell_x}$. This is because the differential operator $-i\tfrac{d}{dy}$ in Eq.~(\ref{H1kxbt}) is associated with the linearization of the periodic Hamiltonian [Eq.~(\ref{H1})] in the vicinity of the Brillouin zone corners with $k_y=0$, while such Brillouin zone corners are projected to $k_x= \tfrac{4\pi}{3\ell_x}, \tfrac{8\pi}{3\ell_x}$, which are located inside our chosen regime but outside the conventional first Brillouin zone. Note that the energy bands in our chosen regime should be the same as those in the conventional first Brillouin zone, even though the ribbon Bloch Hamiltonian [Eq.~(\ref{H1kxbt})] seemingly has a period of $\tfrac{4\pi}{\ell_x}$. In fact, since the real-space period of the ribbon is $\ell_x$ [Fig.~\ref{fig1}(d)], the energy bands must have a period of $\tfrac{2\pi}{\ell_x}$.

In our chosen regime, the guiding center positions should be rewritten as
\begin{align} 
\ell_0^b &=\frac{1}{\lambda_b} \sqrt{\frac{4}{3}\left\{1+\frac{1}{g }\ln \left[ -2\cos \left(\frac{1}{2}k_x\ell_x \right) \right] \right\}^2 -\frac{1}{3}} -\frac{1}{\lambda_b} , \nonumber
\\
\ell_0^t &= \frac{1}{\lambda_t} \sqrt{\frac{4}{3}\left\{1+\frac{1}{g }\ln \left[ -2\cos \left( \frac{1}{2}k_x\ell_x \right) \right] \right\}^2 -\frac{4}{3}},
\end{align}
which can be obtained by setting $k_x\rightarrow k_x+\tfrac{2\pi}{\ell_x}$ in Eq.~(\ref{domain}). In the vicinity of such guiding centers, the ribbon Bloch Hamiltonian [Eq.~(\ref{H1kxbt})] can be linearized to 
\begin{equation} \label{h1}
\begin{split}
\mathcal h_{1,k_x,y}^{b,t}=&\Omega_{b,t}(y-\ell_0^{b,t})\sigma^x - i\gamma_0\ell_y \frac{d}{dy}\sigma^y\tau^z 
\\
&+\frac{1}{2}\gamma_1\tau^x+\frac{1}{2}\gamma_1\sigma^z\tau^x,
\end{split}
\end{equation}
where we have defined $\Omega_{b,t}=- \gamma_0 (\tfrac{d}{dy} \ln |\gamma_0^{b,t}(y)|)_{y=\ell_0^{b,t}}$ for transparency. Remarkably, the ribbon Bloch Hamiltonian [Eq.~(\ref{H1kxbt})] in the real space is now reduced to a coupled Dirac model [Eq.~(\ref{h1})], which can be analytically solved for the dispersions of the strain-induced pseudo Landau levels.

\subsection{Dispersions of pseudo Landau levels}
\label{sec3d}
We now solve the pseudo Landau level dispersions by analytically diagonalizing Eq.~(\ref{h1}). For the tensile strain patterns in consideration, i.e., $\lambda_{b,t}>0$, according to Eq.~(\ref{gamma0bt}), $|\gamma_0^{b,t}(y)|$ is a decreasing function of $y$ such that $\text{sgn} (\Omega_{b,t}) = \text{sgn} (\gamma_0)$. In the realistic Bernal-stacked bilayer graphene, we have $\gamma_0<0$ \cite{jung2014}. As a result, $\mathcal h_{1,k_x,y}^{b,t}$ may be rewritten as
\begin{equation} \label{h1_op}
\mathcal h_{1,k_x,y}^{b,t}=
\begin{bmatrix}
0 & -\epsilon_{b,t} \hat a_{b,t}^\dagger & \gamma_1 & 0
\\
-\epsilon_{b,t} \hat a_{b,t} & 0 & 0 & 0
\\
\gamma_1 & 0 & 0 & -\epsilon_{b,t} \hat a_{b,t}
\\
0 & 0 & -\epsilon_{b,t} \hat a_{b,t}^\dagger & 0
\end{bmatrix},
\end{equation}
where $\hat a_{b,t} = \frac{1}{\sqrt 2}( \xi_{b,t} + \frac{d}{d\xi_{b,t}})$ and $\hat a_{b,t}^\dagger = \frac{1}{\sqrt 2} ( \xi_{b,t} - \frac{d}{d\xi_{b,t}})$ are the quantum oscillator ladder operators with $\xi_{b,t}=(y-\ell_0^{b,t})/\ell_M^{b,t}$ being the dimensionless length scaled by the magnetic length $\ell_M^{b,t}=\sqrt{\gamma_0\ell_y/\Omega_{b,t}}=\sqrt{\hbar v_F/\Omega_{b,t}}$; and $\epsilon_{b,t}=\sqrt{2\Omega_{b,t}\gamma_0\ell_y}=\sqrt 2\hbar v_F/\ell_M^{b,t}$ is the energy scale associated with bend/twist. We dub the eigenvector of the bosonic number operator $\hat a^\dagger \hat a$ as $|n\rangle$, where $n=0,1,2,\cdots$. Such an eigenvector explicitly reads $|n\rangle=(2^n\sqrt \pi n!)^{-1/2} \exp(-\xi_{b,t}^2/2) H_n(\xi_{b,t})$ with $H_n(\cdot)$ being the $n$th Hermite polynomial. In terms of $|n\rangle$, the eigenvector of Eq.~(\ref{h1_op}) may be written as 
\begin{equation} \label{eig_vecn}
\left|\psi_{n\geq2,k_x}^{b,t} \right\rangle=
\begin{bmatrix}
c_{n,k_x,A_l}|n-1\rangle
\\
c_{n,k_x,B_l}|n-2\rangle
\\
c_{n,k_x,A_u}|n-1\rangle
\\
c_{n,k_x,B_u}|n\rangle
\end{bmatrix},
\end{equation}
where we require $n\geq 2$. Substituting Eq.~(\ref{eig_vecn}) into Eq.~(\ref{h1_op}), a nonzero eigenvector requires the following secular equation (applicable for $n\geq 2$)
\begin{equation}
\begin{vmatrix}
\mathcal E_{n,k_x}^{b,t} & \epsilon_{b,t}\sqrt{n-1} & -\gamma_1 & 0
\\
\epsilon_{b,t}\sqrt{n-1} & \mathcal E_{n,k_x}^{b,t} & 0 & 0
\\
-\gamma_1 & 0 & \mathcal E_{n,k_x}^{b,t} &\epsilon_{b,t}\sqrt{n} 
\\
0 & 0 & \epsilon_{b,t}\sqrt{n} & \mathcal E_{n,k_x}^{b,t}
\end{vmatrix}=0,
\end{equation} 
from which the eigenenergy can be resolved as
\begin{align} \label{PLLn}
\mathcal E_{n\geq 2,k_x}^{b,t}= &\pm \sqrt{ \frac{1}{2}\pm \frac{1}{2}\sqrt{1-\frac{4n(n-1)\epsilon_{b,t}^4}{[(2n-1)\epsilon_{b,t}^2+\gamma_1^2]^2}} } \nonumber
\\
&\times \sqrt{(2n-1)\epsilon_{b,t}^2+\gamma_1^2 }.
\end{align}
Equation~(\ref{PLLn}) represents the dispersions of the pseudo Landau levels with $n=2,3,\cdots$, whose  momentum dependence is acquired from the parameter $\epsilon_{b,t}$. The $-$ ($+$) sign under the square root corresponds to the pseudo Landau levels arising from the crossed (gapped) quadratic energy bands [Fig.~\ref{fig1}(b)]. By referring to Eq.~(\ref{eig_vecn}), the eigenvector for the first pseudo Landau level may be written as
\begin{equation} \label{eig_vec1}
\left|\psi_{1,k_x}^{b,t} \right\rangle=
\begin{bmatrix}
c_{1,k_x,A_l}|0\rangle
\\
0
\\
c_{1,k_x,A_u}|0\rangle
\\
c_{1,k_x,B_u}|1\rangle
\end{bmatrix}.
\end{equation}
Plugging Eq.~(\ref{eig_vec1}) into Eq.~(\ref{h1_op}) and requiring nonzero solution for $\{c_{1,k_x,A_l}, c_{1,k_x,A_u}, c_{1,k_x,B_u}\}$, we arrive at the following secular equation
\begin{equation}
\begin{vmatrix}
\mathcal E_{1,k_x}^{b,t}  & -\gamma_1 & 0
\\
-\gamma_1  & \mathcal E_{1,k_x}^{b,t} &\epsilon_{b,t}
\\
0 &  \epsilon_{b,t} & \mathcal E_{1,k_x}^{b,t}
\end{vmatrix}=0.
\end{equation} 
The dispersions of the first pseudo Landau levels can be resolved as
\begin{equation} \label{PLL1}
\mathcal E_{1,k_x}^{b,t}=0, \qquad \mathcal E_{1,k_x}^{b,t}=\pm \sqrt{\epsilon_{b,t}^2+\gamma_1^2},
\end{equation}
which are respectively associated with the first pseudo Landau levels resulting from the two crossed and gapped quadratic energy bands [Fig.~\ref{fig1}(b)]. By drawing an analog to Eq.~(\ref{eig_vecn}) again, the eigenvector of the  zeroth pseudo Landau level may be written as
\begin{equation} \label{eig_vec0}
\left|\psi_{0,k_x}^{b,t} \right\rangle=
\begin{bmatrix}
0
\\
0
\\
0
\\
c_{0,k_x,B_u}|0\rangle
\end{bmatrix},
\end{equation}
from which the dispersion of the zeroth pseudo Landau level is derived as
\begin{equation} \label{PLL0}
\mathcal E_{0,k_x}^{b,t}=0.
\end{equation}
Equations~(\ref{PLLn}),~(\ref{PLL1}), and~(\ref{PLL0}) constitute our key result, analytically characterizing the dispersions of the strain-induced pseudo Landau levels in bent/twisted ribbons of Bernal-stacked bilayer graphene.

\begin{figure*}[t]
\includegraphics[width = \textwidth]{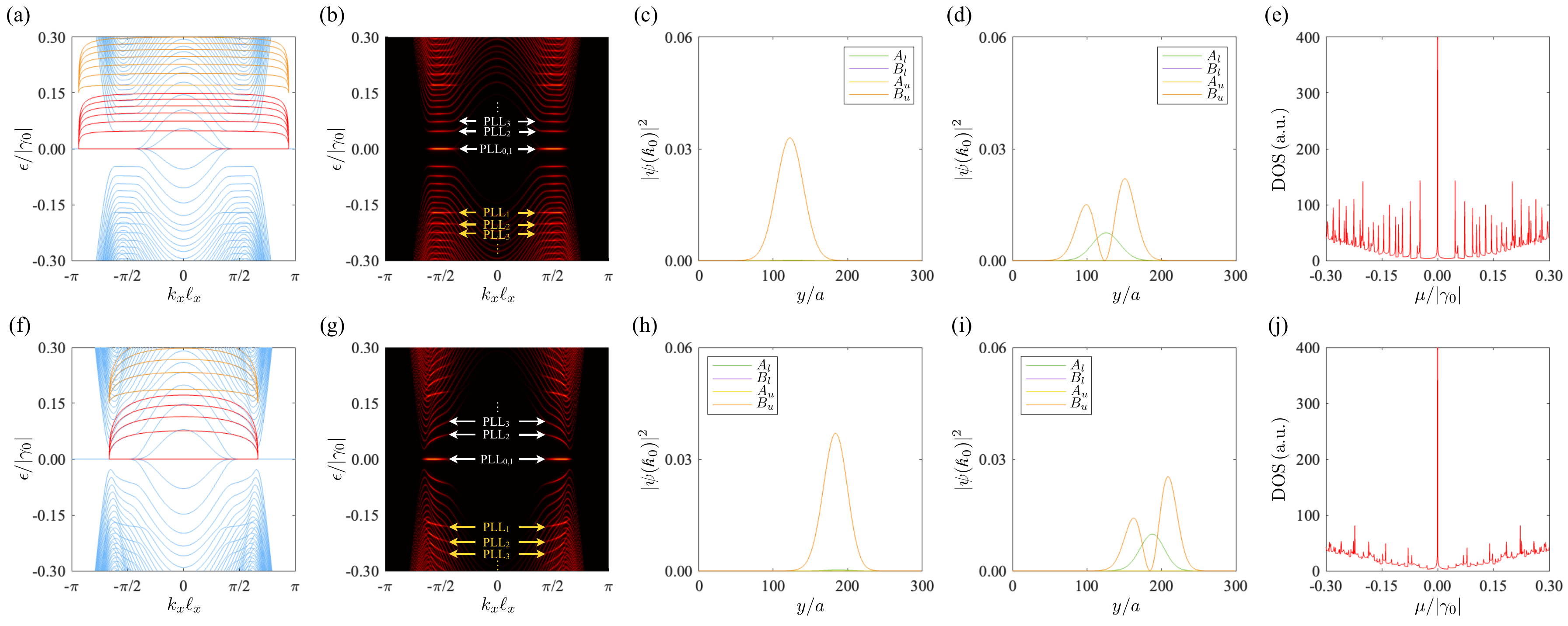}
\caption{(a) Band structure of the bent ribbon of the Bernal-stacked bilayer graphene with $\lambda_b=0.260W^{-1}$. The blue curves are the numerical bands obtained through diagonalization of Eq.~(\ref{H1bt}). The overlaid red (orange) curves are the analytically derived pseudo Landau levels [Eqs.~(\ref{PLLn}),~(\ref{PLL1}), and~(\ref{PLL0})] associated with the two crossed (gapped) energy bands [Fig.~\ref{fig1}(b)]. For clarity, only the lowest few analytic pseudo Landau levels are plotted. (b) The bulk spectral function of panel (a) reveals that those parts of numerical bands, with which the pseudo Landau levels are in accordance, are originated from the bulk. Here the ``bulk'' is defined as the central $50\%$ of the ribbon. The lowest few pseudo Landau levels associated with the two crossed and gapped energy bands [Fig.~\ref{fig1}(b)] are marked with white and yellow arrows, respectively. (c) Sublattice-resolved probability distribution for the zeroth pseudo Landau level in panels (a,b) at a given momentum $\mathcal k_0=1.7\ell_x^{-1}$. The zeroth pseudo Landau level is only hosted by the $B_u$ (orange) sublattice. (d) Sublattice-resolved probability distribution for the zero-energy first pseudo Landau level in panels (a, b) at a given momentum $\mathcal k_0=1.7\ell_x^{-1}$. The first pseudo Landau level is supported by the $A_l$ (green) and $B_u$ (orange) sublattices. (e) Density of states of the bent ribbon. The zeroth and first pseudo Landau levels contribute a peak at the charge neutrality point $\mu=0$. (f-j) are the same as (a-e) except that the strain pattern is twist $\lambda_t=0.766W^{-1}$. In all panels, $\gamma_1=0.15|\gamma_0|$ is used.
} \label{fig3}
\end{figure*}

We now numerically substantiate the analytically derived pseudo Landau level dispersions [Eqs.~(\ref{PLLn}),~(\ref{PLL1}), and~(\ref{PLL0})]. First, we notice the accuracy of the analytic dispersions is justified by the good match to the numerically calculated band structures [Figs.~\ref{fig3}(a, f)]. Second, the spectral functions have revealed the bulk nature of the strain-induced pseudo Landau levels [Figs.~\ref{fig3}(b, g)] and thus justify their origination as the SSH domain-wall modes. Third, the zeroth and first pseudo Landau levels at the charge neutrality point exhibit the correct sublattice polarization. On the one hand, as reflected in the wave function [Eq.~(\ref{eig_vec0})], the zeroth pseudo Landau level only resides on the $B_u$ sublattice [Figs.~\ref{fig3}(c, h)]. On the other hand, even though the wave function of the first pseudo Landau levels  [Eq.~(\ref{eig_vec1})] seems to occupy $A_l$, $A_u$, and $B_u$ sublattices, it is worth noting that the secular equation further requires $c_{1,k_x,A_u}=0$ for $\mathcal E_{1,k_x}^{b,t}=0$, implying that the first pseudo Landau level at the charge neutrality point only resides on the $A_l$ and $B_u$ sublattices [Figs.~\ref{fig3}(d, i)]. The sublattice polarization of the zeroth and first pseudo Landau levels is consistent with the sublattice distribution of the SSH domain-wall modes. 

We have so far justified the analytically derived dispersions of the pseudo Landau levels [Eqs.~(\ref{PLLn}),~(\ref{PLL1}), and~(\ref{PLL0})]. Before leaving the present section, it may be instructive to study these pseudo Landau levels in two limits. On the one hand, for vanishing interlayer hopping $\gamma_1 \rightarrow 0$, the pseudo Landau level dispersions become
\begin{equation}
\left.\mathcal E_{n,k_x}^{b,t}\right|_{\gamma_1\rightarrow 0}=\pm \sqrt{\Omega_{b,t}\gamma_0\ell_y[(2n-1)\pm\text{sgn}(2n-1)]},
\end{equation}
which can be understood as two sets of Dirac-Landau levels arising from the two uncoupled strained graphene monolayers. On the other hand, for strong interlayer coupling $\gamma_1\gg \epsilon_{b,t}$, the pseudo Landau level dispersions are reduced to
\begin{equation}
\begin{split}
\left.\mathcal E_{n,k_x}^{b,t}\right|_{\gamma_1\gg\epsilon_{b,t}} &= \pm \frac{\epsilon_{b,t}^2}{\gamma_1} \sqrt{n(n-1)},
\\
\left.\mathcal E_{n,k_x}^{b,t}\right|_{\gamma_1\gg\epsilon_{b,t}} &=\pm \gamma_1\mp\frac{n(n-1)\epsilon_{b,t}^4}{2\gamma_1^3},
\end{split}
\end{equation}
where the former (latter) characterizes the strain-induced pseudo Landau levels associated with the two crossed (gapped) quadratic energy bands [Fig.~\ref{fig1}(b)]. Intriguingly, the former exhibits the same $n$ dependence as the Landau levels of the Bernal-stacked bilayer graphene subjected to a uniform magnetic field \cite{marcin2011, yin2015} and can capture the flatness of the zeroth and first pseudo Landau levels [Eqs.~(\ref{PLL1}) and~(\ref{PLL0})]. The flatness indicates large density of states [Figs.~\ref{fig3}(e, j)] at the charge neutrality point and essentially results in interaction effects, which are to be detailed in the following section.

\section{Global antiferromagnetism}
\label{sec4}
In Sec.~\ref{sec3}, we have shown that the strain-induced zeroth and first pseudo Landau levels in Bernal-stacked bilayer graphene are flat and sublattice polarized, similar to the zeroth pseudo Landau level in the strained monolayer graphene \cite{settnes2016, shi2021, liu2022}. Moreover, the edge bands of the strained Bernal-stacked bilayer graphene are also flat and sublattice polarized, mimicking their counterparts of the strained monolayer graphene. As the sublattice polarized flat zeroth pseudo Landau level and edge state produce in strained monolayer graphene global antiferromagnetism with edge compensation \cite{roy2014, elias2023}, we thus expect similar interaction-induced magnetism in strained Bernal-stacked bilayer graphene.

At half filling, the density of states of the strained Bernal-stacked bilayer graphene is boosted [Figs.~\ref{fig3}(e, j)] because of the flatness of the zeroth and first pseudo Landau levels and the edge states [Figs.~\ref{fig2}(b, d, f, h) and Figs.~\ref{fig3}(a, b, f, g)]. For simplicity, we model the resulting interaction with an onsite Hubbard term
\begin{equation} \label{H_int}
H_\text{int}= U \sum_{\alpha, x, y}n_{\alpha, x, y,\uparrow} n_{\alpha, x, y, \downarrow},
\end{equation}
where $\alpha=A_l, B_l, A_u, B_u$ is the sublattice index, $(x,y)$ marks the position of the lattice site belonging to the $\alpha$th sublattice, and $n_{\alpha, x, y, \sigma} = c_{\alpha, x, y, \sigma}^\dagger c_{\alpha, x, y, \sigma}$ is the number operator for the spin-$\sigma$ species on that site. At the mean-field level, the Hubbard term is reduced to
\begin{equation} \label{H_int_MF}
H_\text{int}^\text{MF}=U\sum_{\alpha, x,y} \left\langle n_{\alpha, x,y,\downarrow} \right\rangle n_{\alpha, x,y,\uparrow} + \left\langle n_{\alpha, x,y,\uparrow} \right\rangle n_{\alpha, x,y,\downarrow},
\end{equation}
where the energy shift $-U\sum_{\alpha, x,y} \langle n_{\alpha, x,y,\uparrow} \rangle \langle n_{\alpha, x,y,\downarrow} \rangle$ is neglected.  Because of the translational symmetry along the $x$ direction, $\langle n_{\alpha, x, y, \sigma} \rangle$ in principle shows no $x$ dependence, i.e., $\langle n_{\alpha, x, y, \sigma} \rangle=\langle n_{\alpha, y, \sigma} \rangle$. Consequently, Eq.~(\ref{H_int_MF}) is reduced to
\begin{equation} 
H_\text{int}^\text{MF}=U\sum_{\alpha,k_ x,y} \left\langle n_{\alpha,y,\downarrow} \right\rangle n_{\alpha, k_x,y,\uparrow} + \left\langle n_{\alpha,y,\uparrow} \right\rangle n_{\alpha, k_x,y,\downarrow},
\end{equation}
The full interacting problem, $H_1^{b,t}+H_\text{int}^\text{MF}$, can then be solved self-consistently: (i) Initialize $\langle n_{\alpha, y, \sigma} \rangle$, i.e., the spin-$\sigma$ particle number on the site located at $y$ belonging to the $\alpha$th sublattice; (ii) Diagonalize $H_1^{b,t}+H_\text{int}^\text{MF}$ with a unitary transformation $c_{\alpha, k_x, y, \sigma}=\sum_i Q_{\alpha, k_x, \sigma}^{y, i}d_{\alpha, k_x, i, \sigma}$; (iii) Evaluate the resulting particle number according to  $\langle n_{\alpha, y, \sigma} \rangle = \tilde N^{-1} \sum_{p_x}\sum_i^\text{occ} (Q_{\alpha, p_x, \sigma}^{i,y})^* Q_{\alpha, p_x, \sigma}^{y,i}$, where the sum of $i$ runs over the occupied eigenstates; and (iv) Repeating (i)-(iii) until the value of $\langle n_{\alpha, y, \sigma} \rangle$ converges up to an accuracy of 1\%. Then, the net spin distribution for the $\alpha$th sublattice reads
\begin{equation}
S_\alpha(y)= \bar n_{\alpha, y, \uparrow}-\bar n_{\alpha, y, \downarrow},
\end{equation} 
where $\bar n_{\alpha, y, \sigma}$ is the converged value of $\langle n_{\alpha, y, \sigma} \rangle$. We mention that the allowed $y$ coordinates depend on the sublattice species, i.e., $y_{A_{l,i}}=y_{A_{u,i}}=y_{B_{l,i}}-\tfrac1 2 a=y_{B_{u,i}}+\tfrac 1 2 a$ [see Fig.~\ref{fig1}(d)]. In the following, the coordinate difference between any two sublattices, if any, is neglected for simplicity. Consequently, we may define the following local ferromagnetic and antiferromagnetic order parameters
\begin{equation}
\begin{split}
M(y)=S_{A_l}(y)+S_{B_l}(y)+S_{A_u}(y)+S_{B_u}(y),
\\
N(y)=S_{A_l}(y)-S_{B_l}(y)-S_{A_u}(y)+S_{B_u}(y),
\end{split}
\end{equation}
where the former represents the net spins at a give position, while the latter measures the net spin bias among four sublattices at the same position. 

\begin{figure}[t]
\includegraphics[width = 8.6cm]{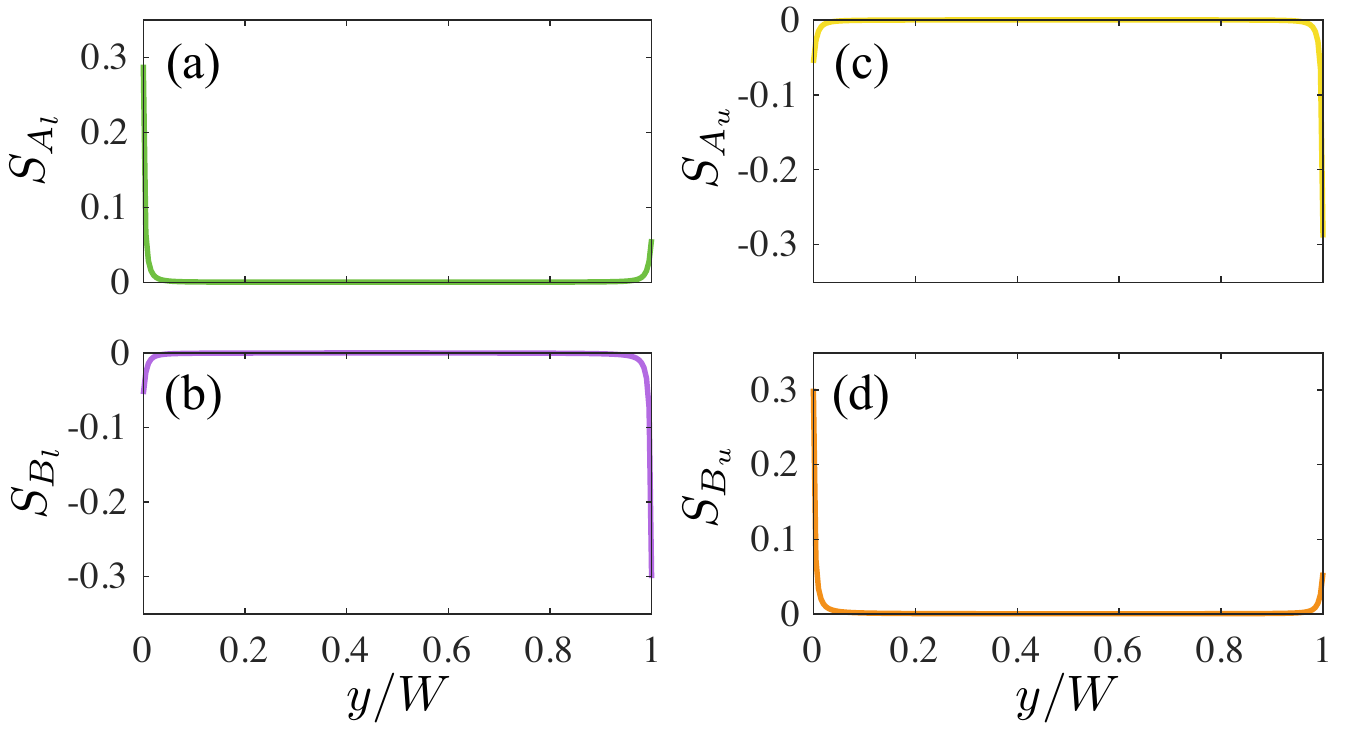}
\caption{Net spin distributions on the $A_l$ (green), $B_l$ (purple), $A_u$ (yellow), and $B_u$ (orange) sublattices of a strain-free ribbon of Bernal-stacked bilayer graphene with Hubbard interaction. In all panels, $\gamma_1=0.15|\gamma_0|$ and $U=1.2|\gamma_0|$ are used.
} \label{fig4}
\end{figure}

\begin{figure}[b]
\includegraphics[width = 8.6cm]{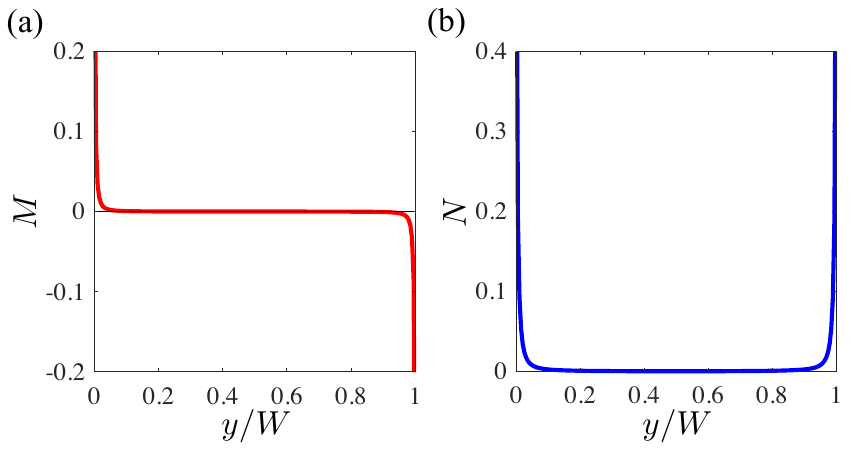}
\caption{Magnetism of a strain-free ribbon of Bernal-stacked bilayer graphene with Hubbard interaction. (a) Local ferromagnetic order parameter. (b) Local antiferromagnetic order parameter. In both panels, $\gamma_1=0.15|\gamma_0|$ and $U=1.2|\gamma_0|$ are used.
} \label{fig5}
\end{figure}

We first study the net spin distributions and the resulting magnetism in a strain-free ribbon of Bernal-stacked bilayer graphene, which hosts sublattice-polarized flat topological edge bands. We observe that net spins tend to accumulate at the four edges: net up-spins at the $y=0$ edge of the lower monolayer terminated at the $A_l$ sublattice [Fig.~\ref{fig4}(a)]; net down-spins at the $y=W$ edge of the lower monolayer terminated at the $B_l$ sublattice [Fig.~\ref{fig4}(b)]; net down-spins at the $y=W$ edge of the upper monolayer terminated at the $A_u$ sublattice [Fig.~\ref{fig4}(c)]; and net up-spins at the $y=0$ edge of the upper monolayer terminated at the $B_u$ sublattice [Fig.~\ref{fig4}(d)]. Such sublattice-polarized net spin distributions are consistent with the sublattice polarization of the associated flat topological edge bands. We note that each of the four curves in Figs.~\ref{fig4}(a-d) also exhibits a small tail, which may arise from the decay of the topological edge states supported by the adjacent sublattices. As the decay length of the topological edge states is typically much smaller than the ribbon width $W$, no net spins are expected in the bulk [Figs.~\ref{fig4}(a-d)]. In the view of magnetism, we find the net spin distributions result in opposite (identical) local ferromagnetic (antiferromagnetic) orders at the edges while there is no magnetism in the bulk [Figs.~\ref{fig5}(a, b)]. Such characters of magnetism resemble those of a monolayer graphene ribbon \cite{hikihara2003, lee2005, son2006, feldner2010, jung2010, jung2011, yazyev2011, hu2012, golor2013, magda2014}. The interlayer antiferromagnetism has also been verified by means of quantum Monte Carlo simulations and functional renormalization group calculations \cite{lang2012, scherer2012}

\begin{figure*}[t]
\includegraphics[width = \textwidth]{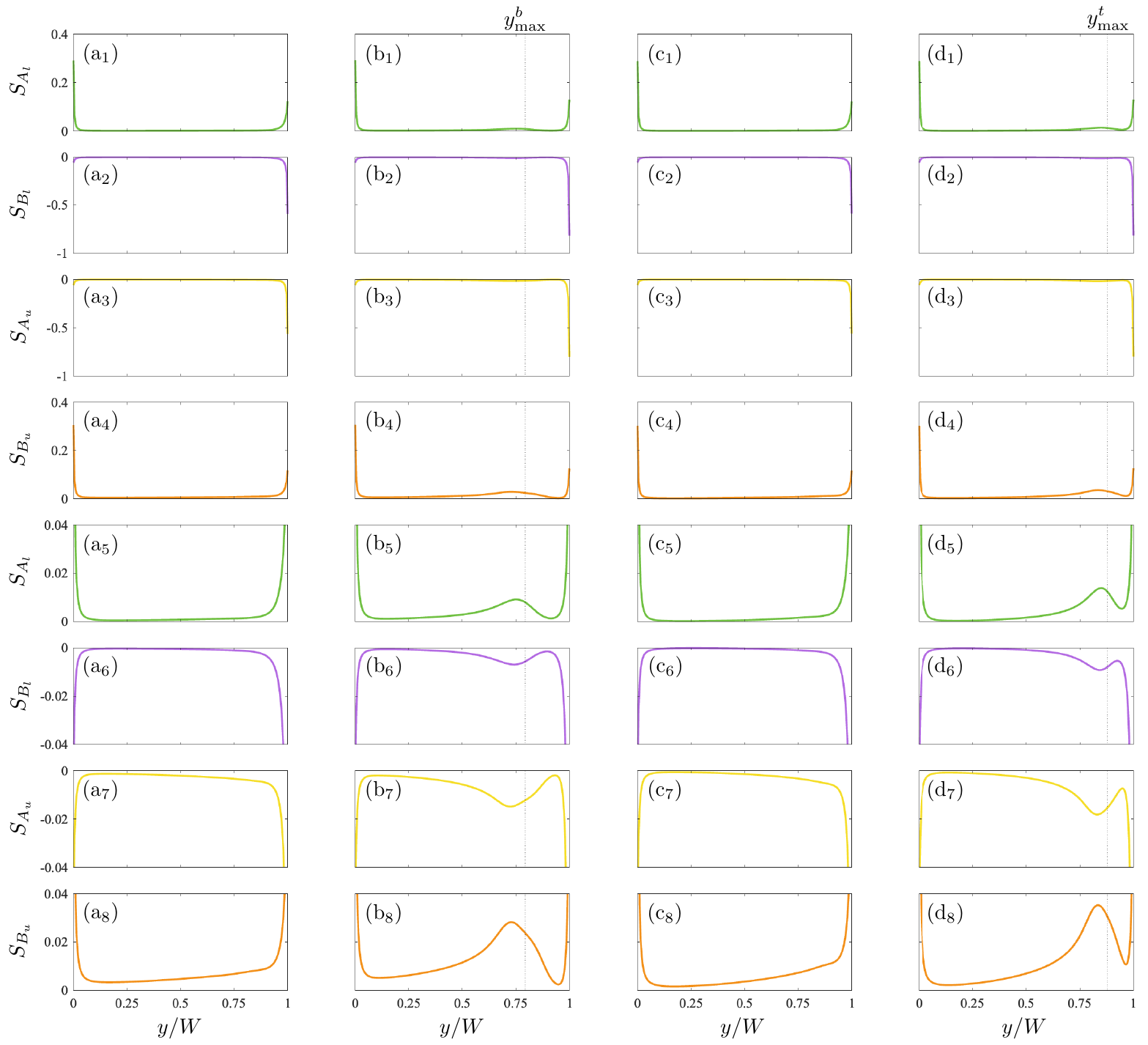}
\caption{Net spin distributions on the $A_l$ (green), $B_l$ (purple), $A_u$ (yellow), and $B_u$ (orange) sublattices of a strained ribbon of Bernal-stacked bilayer graphene with Hubbard interaction. (a$_1$-a$_8$) Bend $\lambda_b=0.209W^{-1}$ with panels (a$_1$-a$_4$) showing the full net spin distributions on each sublattice and (a$_5$-a$_8$) being the enlargements of panels (a$_1$-a$_4$). (b$_1$-b$_8$) are similar to (a$_1$-a$_8$) but for a stronger bend $\lambda_b=0.336W^{-1}$. The dotted lines mark the top of the phase boundary $y_{\max}^b=0.794W$ [Fig.~\ref{fig2}(c)]. (c$_1$-c$_8$) are similar to (a$_1$-a$_8$) but for twist $\lambda_t=0.679W^{-1}$. (d$_1$-d$_8$) are also similar to (a$_1$-a$_8$) but for a stronger twist $\lambda_t=0.885W^{-1}$. The dotted lines mark the top of the phase boundary $y_{\max}^t=0.879W$ [Fig.~\ref{fig2}(g)]. In all panels,  $\gamma_1=0.15|\gamma_0|$ and $U=1.2|\gamma_0|$ are used.
} \label{fig6}
\end{figure*}

\begin{figure}[t]
\includegraphics[width = 8.6cm]{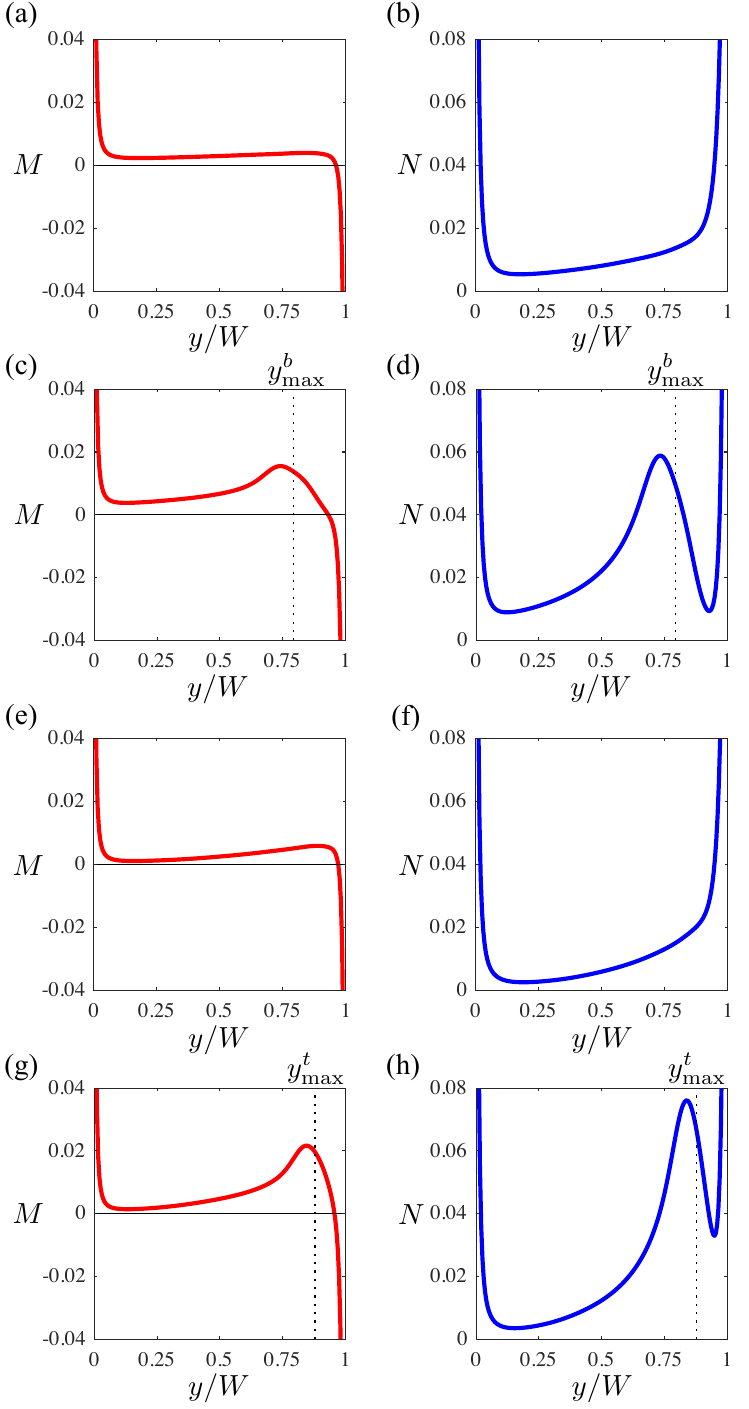}
\caption{Magnetism of strained ribbons of Bernal-stacked bilayer graphene with Hubbard interaction. (a, c, e, g) Local ferromagnetic order parameters for various strained ribbons. (b, d, f, h) Local antiferromagnetic order parameters for various strained ribbons. (a, b) Bend $\lambda_b=0.209W^{-1}$. (c, d) Bend $\lambda_b=0.336W^{-1}$. (e, f) Twist $\lambda_t=0.679W^{-1}$. (g, h) Twist $\lambda_t=0.885W^{-1}$. In all panels, $\gamma_1=0.15|\gamma_0|$ and $U=1.2|\gamma_0|$ are used.
} \label{fig7}
\end{figure}

We then turn to investigate the net spin distributions and the resulting magnetism in bent ribbons of Bernal-stacked bilayer graphene, which host sublattice-polarized flat topological edge bands and pseudo Landau levels. For weak bend $\lambda_b=0.209W^{-1}<\lambda_b^\text{c}$, we first note that the net spin distributions around the $y=0$ edge are almost intact comparing to their strain-free counterparts, while the net spin distributions around the $y=W$ edge are enhanced [Figs.~\ref{fig6}(a$_1$-a$_4$) versus Figs.~\ref{fig4}(a-d)]. This is because the local strain at the $y=0$ edge is approximately zero, while the local strain at the $y=W$ edge is increased resulting in elongated flat topological edge bands in the regime $\mathcal k_{\min}^b\leq |k_x| \leq \mathcal k_{\max}^b$ [Figs.~\ref{fig2}(a, b)], which can potentially host more net spins under interaction. Besides the elongated flat topological edge bands, there are zeroth and first pseudo Landau levels in the same regime (see Sec.~\ref{sec3b}). Similar to the topological edge bands, these flat pseudo Landau levels also support net spin distributions but in the bulk [Figs.~\ref{fig6}(a$_5$-a$_8$)], in contrast to the zero bulk net spin distributions in the strain-free ribbons [Figs.~\ref{fig4}(a-d)]. As the zeroth (first) pseudo Landau level resides on the $B_u$ sublattice ($A_l$ and $B_u$ sublattices) (see Sec.~\ref{sec3d}), net spins should in principle only reside on the $A_l$ and $B_u$ sublattices in the bulk [Figs.~\ref{fig6}(a$_5$, a$_8$)]. Interestingly, one also observes in the bulk net spin distributions on the $B_l$ and $A_u$ sublattices [Figs.~\ref{fig6}(a$_6$, a$_7$)], where the zeroth and first pseudo Landau levels do not have supports. Such net spin distributions may arise from the decay of the zeroth and first pseudo Landau levels hosted by the $A_l$ and $B_u$ sublattices. The net spin distributions of a bent ribbon produce  local ferromagnetic and antiferromagnetic orders [Figs.~\ref{fig7}(a, b)]. The local ferromagnetic order parameters are of opposite signs at opposite edges [Fig.~\ref{fig7}(a)], similar to the strain-free case [Fig.~\ref{fig5}(a)]. However, in contrast to the strain-free case, the edge ferromagnetic order parameters do not cancel because of the ferromagnetism in the bulk [Fig.~\ref{fig7}(a)]. As for the local antiferromagnetic order parameter, the two edges are now unbalanced and the bulk is also antiferromagnetic [Fig.~\ref{fig7}(b)], distinguished from the strain-free case [Fig.~\ref{fig5}(b)]. Such local ferromagnetic and antiferromagnetic orders resemble those in the strained monolayer graphene \cite{roy2014, elias2023}, which exhibits global antiferromagnetism with edge compensation, and can thus be understood as the global antiferromagnetism of the strained bilayer graphene. For strong bend $\lambda_b=0.336W^{-1}>\lambda_b^\text{c}$, we find that the net spin distributions and magnetism are similar to those of the weak bend case except that both features exhibit peaks in the bulk [Figs.~\ref{fig6}(b$_1$-b$_8$) and~\ref{fig7}(c, d)]. The emergence of these peaks is consistent with the fact that the guiding center of the pseudo Landau levels is bounded from $y_{\max}^b$ [Fig.~\ref{fig2}(c)]. The peaks sightly deviate from $y_{\max}^b$, because the pseudo Landau levels are not perfectly localized. Such peaks imply more pronounced local magnetic orders in the bulk and indicate that the global antiferromagnetism becomes stronger with increased bend.

Lastly, we briefly mention the net spin distributions and the resulting magnetism in twisted ribbons of Bernal-stacked bilayer graphene, which host sublattice-polarized flat topological edge bands and pseudo Landau levels. For weak twist $\lambda_t=0.679W^{-1}<\lambda_t^\text{c}$, the net spin distributions and the magnetism in the vicinity of edges [Figs.~\ref{fig6}(c$_1$-c$_8$) and~\ref{fig7}(e, f)] resemble those of the weak bend $\lambda_b=0.209W^{-1}$ [Figs.~\ref{fig6}(a$_1$-a$_8$) and~\ref{fig7}(a, b)], because the two strain patterns are approximately the same around the edges. On the other hand, the curves of the bulk net spin distributions and magnetism of the twisted ribbon [Figs.~\ref{fig6}(c$_1$-c$_8$) and~\ref{fig7}(e, f)] are steeper than those of the bent ribbon [Figs.~\ref{fig6}(a$_1$-a$_8$) and~\ref{fig7}(a, b)]. This is likely because there are more (less) allowed pseudo Landau level guiding centers located near the $y=W$ ($y=0$) edge in the twisted ribbon than in the bent ribbon [Fig.~\ref{fig2}(e) versus Fig.~\ref{fig2}(a)]. For strong twist $\lambda_t=0.885W^{-1}>\lambda_t^\text{c}$, the net spin distributions and magnetism are similar to those of the weak twist case except for the peaks in the bulk [Figs.~\ref{fig6}(d$_1$-d$_8$) and~\ref{fig7}(g, h)], analogous to what happens in bent ribbons. Comparing to the strong bend case, the curves of bulk net spin distributions and magnetism are steeper as expected.

\section{Conclusions}
\label{sec5}
In this paper, we analytically derive the pseudo Landau level dispersions for both bent and twisted ribbons of Bernal-stacked bilayer graphene and show that the Hubbard interaction induces global antiferromagnetism in such ribbons. 

We first briefly review the models of bulk and ribbons of the strain-free Bernal-stacked bilayer graphene. For the latter, the unit cell becomes a two-legged Su-Schrieffer-Heeger model, whose topological end modes realize sublattice-polarized edge states.

We then study the electronic structures of bent and twisted ribbons of Bernal-stacked bilayer graphene. Bend and twist are incorporated by modulating the hopping parameters. We find that the ribbon unit cell becomes a two-legged Su-Schrieffer-Heeger model with a domain wall. According to the band topology analysis, the domain wall turns out to be the guiding center of the strain-induced pseudo Landau levels. Consequently, for the purpose of studying pseudo Landau levels, it is sufficient to reduce the full model in the vicinity of the domain wall. Remarkably, the reduced model is an exactly solvable coupled Dirac model, giving rise to dispersive pseudo Landau levels. In particular, the zeroth and first pseudo Landau levels are flat and sublattice-polarized, analogous to the topological edge states.

Lastly, we investigate the interaction-induced net spin distributions and magnetism in ribbons of Bernal-stacked bilayer graphene. For the strain-free ribbons, net spins turn to accumulate at the edges and result in local ferromagnetic and antiferromagnetic orders there. In contrast, for both bent and twisted ribbons, net spin distributions are also supported in the bulk by the zeroth and first pseudo Landau levels. Specifically, peak features show up if the strain inhomogeneity surpasses a threshold, indicating the bound of the guiding center of the pseudo Landau levels. The net spin distributions associated with the zeroth and first pseudo Landau levels produce bulk ferromagnetism. Combining the edge ferromagnetism, a global antiferromagnetic order emerges and is a direct generalization of the global antiferromagnetism of the strained monolayer graphene.

Our approach of deriving pseudo Landau level dispersions are directly transplantable to Dirac matter under nonuniform magnetic fields. It should also help accurately quantify the dispersive pseudo Landau levels proposed in strained $d$-wave superconductors \cite{nica2018, massarelli2017} and kagome crystals \cite{liu2020}. It would be of importance to check for the strain-induced pseudo Landau levels other interaction effects beyond the global antiferromagnetism, such as superconductivity and the fractional quantum anomalous Hall effect. As the pseudo Landau levels are the domain-wall modes of the two-legged Su-Schrieffer-Heeger chain in nature, they can possibly exhibit non-Abelian statistics \cite{wuyijia2020}, whose strain tunability deserves a careful study.

\begin{acknowledgments}
The authors are indebted to R. Moessner, Jihang Zhu, I. Khaymovich, Junsong Sun, M. Franz, and A. M. Cook. This work is supported by the National Key R\&D Program of China (2022YFA1403700), Innovation Program for Quantum Science and Technology (2021ZD0302400), the National Natural Science Foundation of China (12304196, 12304177, 12350402, 12074022, and 11925402), Guangdong Basic and Applied Basic Research Foundation (2022A1515111034, 2023B0303000011), Guangdong Provincial Quantum Science Strategic Initiative (GDZX2201001, GDZX2401001), Guangdong province (2020KCXTD001), the Natural Science Foundation of Jiangsu Province (BK20230907), the Science, Technology and Innovation Commission of Shenzhen Municipality (ZDSYS20190902092905285), and Center for Computational Science and Engineering of SUSTech. T.L. gratefully acknowledges the scholarship from the Max-Planck-Gesellschaft, by which this project is partially supported.
\end{acknowledgments}

\appendix
\section{Realistic Bernal-stacked bilayer graphene}
\label{app_a}
In Sec.~\ref{sec2} of the main text, we have analyzed the band structure of the strain-free Bernal-stacked bilayer graphene with a simplified tight-binding model enclosing only the nearest-neighbor hoppings. To better characterize the strain-free Bernal-stacked bilayer graphene, more hoppings had better be incorporated into the tight-binding models \cite{jung2014}. Amongst them, the second-nearest-neighbor interlayer hoppings turn out to be the most dominant contribution \cite{partoens2006, min2007, malard2007, zhang2008, gruneis2008, nilsson2008, gava2009, kuzmenko2009, jung2014}. In this section, the second-nearest-neighbor interlayer hoppings are also considered to construct a more realistic model of Bernal-stacked bilayer graphene.

\begin{figure*}[t]
\includegraphics[width = \textwidth]{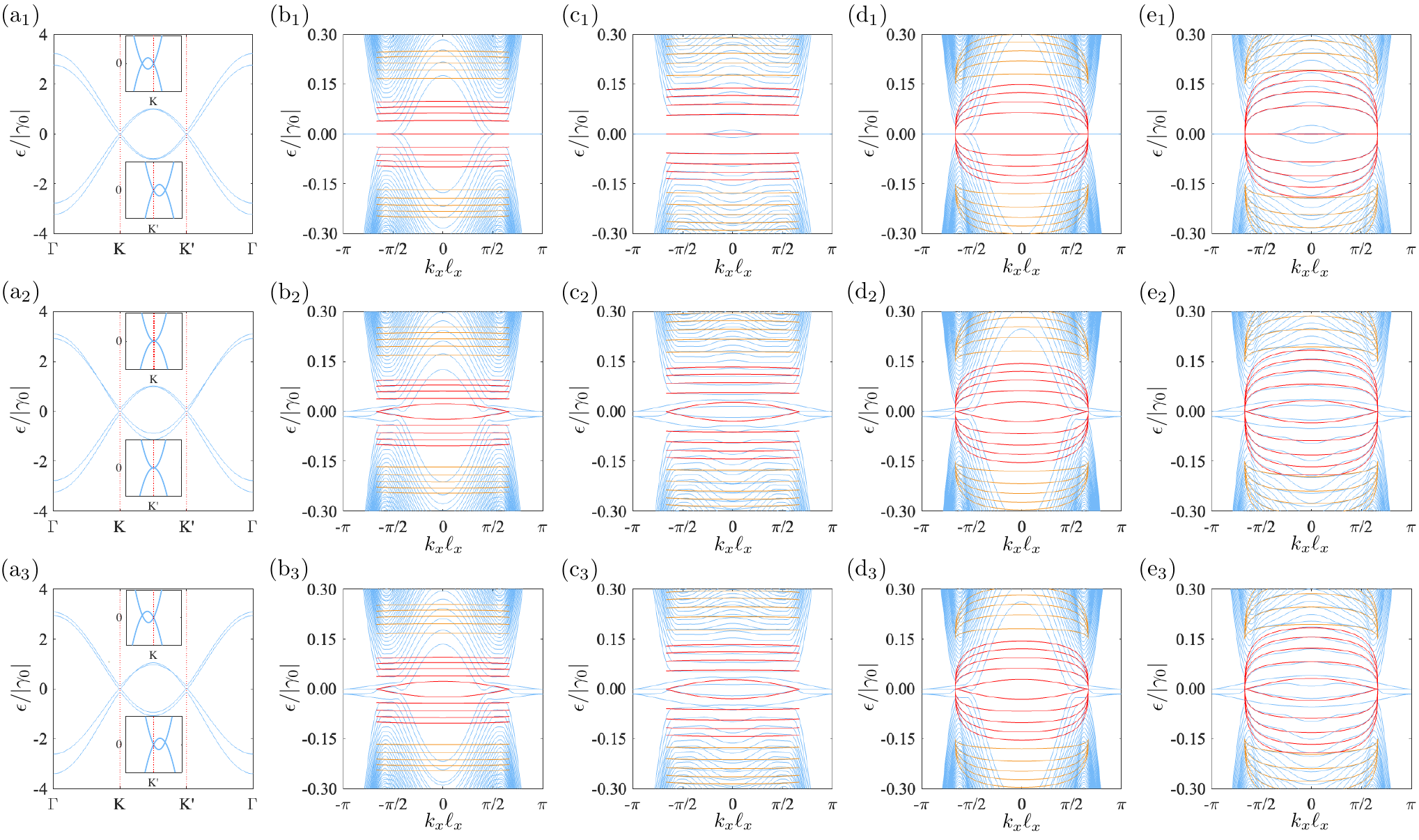}
\caption{Band structures (blue) for the realistic Bernal-stacked bilayer graphene numerically obtained by diagonalization. (a$_1$-e$_1$) Spectra in the presence of second-nearest-neighbor interlayer warping, i.e, $\gamma_3\neq0$ but $\gamma_4=0$. (a$_2$-e$_2$) Spectra in the presence of second-nearest-neighbor interlayer coupling, i.e, $\gamma_3=0$ but $\gamma_4\neq0$. (a$_3$-e$_3$) Spectra in the presence of both types of second-nearest-neighbor interlayer hoppings, i.e, $\gamma_3\neq0$ and $\gamma_4\neq0$. (a$_1$-a$_3$) Bulk bands of the realistic Bernal-stacked bilayer graphene plotted along the high-symmetry path [see inset of Fig.~\ref{fig1}(b)]. The insets in each panel respectively zoom in the bands in the vicinity of the $K$ and $K'$ points. (b$_1$-b$_3$) Spectra of a bent ribbon of the realistic Bernal-stacked bilayer graphene with $\lambda_b=0.209W^{-1}$. The red and orange curves overlaid are the  pseudo Landau levels respectively associated with the quadratically crossed and gapped energy bands [Fig.~\ref{fig1}(b)]. (c$_1$-c$_3$) are similar to (b$_1$-b$_3$) except that the bend is $\lambda_b=0.336W^{-1}$. (d$_1$-d$_3$) are similar to (b$_1$-b$_3$) except for the strain pattern being twist with $\lambda_t=0.679W^{-1}$. (e$_1$-e$_3$) are also similar to (b$_1$-b$_3$) except for the strain pattern being twist with $\lambda_t=0.885W^{-1}$. The parameters used are $\gamma_1=0.138|\gamma_0|$, $\gamma_3=0.108|\gamma_0|$, and $\gamma_4=0.053|\gamma_0|$, taken from Ref.~\cite{jung2014}.
} \label{fig8}
\end{figure*}

We first consider the second-nearest-neighbor hopping between two non-dimer sites $B_l$ and $B_u$, labeled as $\gamma_3$ [Fig.~\ref{fig1}(a)].  Such an interlayer hopping produces trigonal warping in the vicinity of the Brillouin zone corners \cite{mccann2006, mccann2013}. Explicitly, the warping is characterized by
\begin{equation} \label{H2warp}
H_2^{\text{warp}}= \sum_{\bm r, j} \gamma_3 (b_{l,\bm r}^\dagger b_{u,\bm r+\bm \alpha_j} + b_{u,\bm r+\bm \alpha_j} ^\dagger b_{l,\bm r}),
\end{equation}
which, upon Fourier transform, yields in the Bloch Hamiltonian an extra term
\begin{equation}
\begin{split}
\mathcal H_{2,\bm k}^{\text{warp}}=\tfrac{1}{2} \gamma_3 \Re(f_{\bm k}) \tau^x - \tfrac{1}{2} \gamma_3 \Re(f_{\bm k}) \sigma^z \tau^x 
\\
- \tfrac{1}{2} \gamma_3 \Im(f_{\bm k}) \tau^y + \tfrac{1}{2} \gamma_3 \Im(f_{\bm k}) \sigma^z \tau^y.
\end{split}
\end{equation}
The electronic structure of the Bernal-stacked bilayer graphene with trigonal warping (i.e., the spectrum of $\mathcal H_{1,\bm k}+\mathcal H_{2,\bm k}^{\text{warp}}$) then becomes
\begin{equation} \label{E2warp}
\epsilon_{\bm k}=\xi \sqrt{\gamma_0^2|f_{\bm k}|^2 +\frac{1}{2}(\gamma_1^2+\gamma_3^2|f_{\bm k}|^2) + \eta \varepsilon_{\bm k}^2 },
\end{equation}
where we define $\varepsilon_{\bm k}^2= [\tfrac 1 4 (\gamma_1^2-\gamma_3^2|f_{\bm k}|^2)^2+ \gamma_0^2\gamma_3^2|f_{\bm k}|^4 + 2\gamma_0^2\gamma_1\gamma_3 |f_{\bm k}|^3\cos(3\theta_{\bm k})+\gamma_0^2\gamma_1^2|f_{\bm k}|^2]^{1/2}$ with $\theta_{\bm k}=\arg(f_{\bm k})$. The spectrum is illustrated in Fig.~\ref{fig8}(a). In the presence of the trigonal warping, the quadratic band touching at a Brillouin zone corner is broken into four Dirac points, which are respectively associated with $f_{\bm k}=0,\tfrac{\gamma_1\gamma_3}{\gamma_0^2}, \tfrac{\gamma_1\gamma_3}{\gamma_0^2} \exp(i\tfrac{2\pi}{3}), \tfrac{\gamma_1\gamma_3}{\gamma_0^2} \exp(i\tfrac{4\pi}{3})$. The one associated with $f_{\bm k}=0$ is located right at the Brillouin zone corner, while the other three are $C_{3}$-symmetrically located around the Brillouin zone corner with a distance $\tfrac{2\gamma_1\gamma_3}{3a\gamma_0^2}$.

We then consider the effect of second-nearest-neighbor hoping between a dimer site $A_l$ ($A_u$) and a non-dimer site $B_u$ ($B_l$), labeled as $\gamma_4$ [Fig.~\ref{fig1}(a)]. Such an interlayer hopping is characterized by
\begin{equation} \label{H2coup}
H_2^{\text{coup}}=\sum_{\bm r,j} \gamma_4 (a_{l,\bm r}^\dagger b_{u,\bm r-\bm \alpha_j} + a_{u,\bm r}^\dagger b_{l,\bm r+\bm \alpha_j} + \text{H.c.}),
\end{equation}
which, upon Fourier transform, brings to the Bloch Hamiltonian an additional term
\begin{equation}
\mathcal H_{2,\bm k}^{\text{coup}}=\gamma_4\Re(f_{\bm k})\sigma^x\tau^x + \gamma_4 \Im(f_{\bm k}) \sigma^x\tau^y.
\end{equation}
The electronic structure of the Bernal-stacked bilayer graphene with interlayer coupling between a dimer site and a non-dimer site (i.e., the spectrum of $\mathcal H_{1,\bm k}+\mathcal H_{2,\bm k}^{\text{coup}}$) then becomes
\begin{equation} \label{E2coup}
\epsilon_{\bm k} = \eta \sqrt{(\gamma_0+\xi\gamma_4)^2|f_{\bm k}|^2+\frac{\gamma_1^2}{4}} + \xi\frac{\gamma_1}{2},
\end{equation}
which differs from Eq.~(\ref{E1}) by a substitution $\gamma_0 \rightarrow \gamma_0+\xi\gamma_4$. In the vicinity of the Brillouin zone corners, the spectrum is reduced to $\epsilon_{\bm k_x+\bm q}=(\eta+\xi)\tfrac{\gamma_1}{2}+\eta\hbar^2\tilde{v}_F^2 q^2/\gamma_1$, where $\tilde{v}_F=\tfrac{3a}{2\hbar}|\gamma_0+\xi\gamma_4|$ further results in a change in the effective mass. It is worth noting that the quadratic band bottom, touching, and top are immune to this change and are still at $\epsilon=\gamma_1, 0, -\gamma_1$, respectively [Fig.~\ref{fig8}(b)].

The realistic model of Bernal-stacked bilayer graphene simultaneously incorporates both of the above second-nearest-neighbor hoppings and it reads $\mathcal H_{1,\bm k}+\mathcal H_{2,\bm k}^{\text{warp}}+\mathcal H_{2,\bm k}^{\text{coup}}$. Unfortunately, the electronic structure of such a model cannot be written in a simple analytic form. For this reason, we here only present the numeric spectrum, from which both Dirac points arising from trigonal warping and change in effective mass resulting from coupling can be seen [Fig.~\ref{fig8}(c)].

\section{Strain-induced pseudo Landau levels in realistic Bernal-stacked bilayer graphene}
\label{app_b}
In Appendix~\ref{app_a}, we have constructed a more realistic model of Bernal-stacked bilayer graphene by including the second-nearest-neighbor interlayer hoppings. It would be instructive to study the strain effects in this model. As $\gamma_3,\gamma_4 < \gamma_1 \ll \gamma_0$, we here investigate how the additional second-nearest-neighbor interlayer hoppings change the dispersions of the pseudo Landau levels with a perturbative approach.

For the strain patterns in consideration, i.e., bend and twist, the second-nearest-neighbor interlayer hoppings are spatially modulated in a similar way to Eq.~(\ref{sub}) as
\begin{equation} \label{sub34}
\gamma_{3,4}(\bm r,\bm r')=\gamma_{3,4}\exp\left(-g\frac{|\bm r'-\bm r|-\sqrt{a^2+d^2}}{\sqrt{a^2+d^2}} \right),
\end{equation}
where $\bm r$ and $\bm r'$ mark the positions of the second-nearest neighbors across the bilayer and $d=2.4a$ is the interlayer spacing \cite{mccann2013}. In Eq.~(\ref{sub34}), the strain-modulated spacing $|\bm r-\bm r'|$ is found to be $[\alpha_{i,x}^2 (1+\lambda_b y)^2+\alpha_{i,y}^2+d^2]^{1/2}$ for the circular bend and $[(1+\lambda_b^2 y^2)\alpha_{i,x}^2+\alpha_{i,y}^2+d^2]^{1/2}$ for the twist. Therefore, the second-nearest-neighbor interlayer hoppings associated with $\pm\bm \alpha_3$ are still $\gamma_{3,4}$, while those associated with $\pm\bm\alpha_{1,2}$ for the circular bend and the twist are respectively
\begin{align} \label{gamma34bt}
\gamma_{3,4}^b(y) &=\gamma_{3,4} \exp \left[ g\left( 1-\sqrt{\frac{1+\frac 3 4 \lambda_b^2 y^2 + \frac{3}{2}\lambda_b y+\frac{d^2}{a^2}}{1+\frac{d^2}{a^2}}} \right) \right], \nonumber
\\
\gamma_{3,4}^t(y) &=\gamma_{3,4} \exp \left[ g\left( 1-\sqrt{\frac{1+\frac 3 4 \lambda_t^2 y^2+\frac{d^2}{a^2}}{1+\frac{d^2}{a^2}}} \right) \right].
\end{align}
For the strain patterns in consideration, we have $\frac 3 4 \lambda_b^2 y^2 + \frac{3}{2}\lambda_b y, \frac 3 4 \lambda_t^2 y^2 \ll 1+\frac{d^2}{a^2}$, implying $\gamma_{3,4}^{b,t}(y)\lesssim\gamma_{3,4}$. As the second-nearest-neighbor terms are already perturbative, the difference between the strained-modulated and bare hoppings can be safely neglected. Hence, we take $\gamma_{3,4}^{b,t}(y)\approx\gamma_{3,4}$ for simplicity in the following.

We first consider the strain-induced pseudo Landau levels in the presence of second-nearest-neighbor interlayer hoppings between non-dimer sites. As we have made the approximation $\gamma_{3}^{b,t}(y)\approx\gamma_{3}$, this interlayer warping in the presence of strain is still characterized by Eq.~(\ref{H2warp}). Perform Fourier transform along the $x$ direction, Eq.~(\ref{H2warp}) becomes
\begin{equation} \label{H2warp_rib}
\begin{split}
H_2^{\text{warp}}= &\sum_{k_x, y} b_{l,k_x,y}^\dagger \left[2\gamma_3 \cos\left(\frac 1 2 k_x\ell_x\right) + \gamma_3 \hat s_{-\ell_y}\right] b_{u,k_x,y+\frac{\ell_y}{3}} 
\\
+ &\sum_{k_x, y} b_{u,k_x,y+\frac{\ell_y}{3}}^\dagger \left[2\gamma_3 \cos\left(\frac 1 2 k_x\ell_x\right) + \gamma_3 \hat s_{\ell_y}\right] b_{l,k_x,y}.
\end{split}
\end{equation}
In the continuum limit, $\hat s_{\pm \ell_y}=1\pm\ell_y\frac{d}{dy}=1\pm\frac{\epsilon_{b,t}}{2\gamma_0}(\hat a_{b,t}-\hat a_{b,t}^\dagger)$, where we have used $\hat a_{b,t}=\tfrac {1} {\sqrt{2}}(\xi_{b,t}+\tfrac{d}{d\xi_{b,t}})$, $\hat a_{b,t}^\dagger=\tfrac {1} {\sqrt{2}}(\xi_{b,t}-\tfrac{d}{d\xi_{b,t}})$, $\xi_{b,t}=(y-\ell_0^{b,t})/\ell_M^{b,t}$, $\ell_M^{b,t}=\sqrt{\gamma_0\ell_y/\Omega_{b,t}}$, and $\epsilon_{b,t}=\sqrt{2\Omega_{b,t}\gamma_0\ell_y}$. The kernel of Eq.~(\ref{H2warp_rib}) then becomes
\begin{equation} \label{H2kxwarp}
\mathcal {H}_{2,k_x,y}^{\text{warp},b,t}=
\begin{bmatrix}
0 & 0 & 0 & 0
\\
0 & 0 & 0 & \mathcal h_{k_x,-}^{\text{warp}, b,t}
\\
0 & 0 & 0 & 0
\\
0 & \mathcal h_{k_x,+}^{\text{warp},b,t} & 0 & 0
\end{bmatrix},
\end{equation}
where $\mathcal h_{k_x,\pm}^{\text{warp},b,t} =2\gamma_3\cos(\tfrac{1}{2}k_x\ell_x)+\gamma_3 \pm \tfrac{\gamma_3}{2\gamma_0}\epsilon_{b,t}(\hat a_{b,t}-\hat a_{b,t}^\dagger)$ is defined for transparency. Treating $\mathcal {H}_{2,k_x,y}^{\text{warp},b,t}$ as perturbation, the first-order energy corrections to the non-degenerate pseudo Landau levels with  $n\geq2$ are
\begin{equation} \label{dE_warpn}
 \Delta\mathcal E_{n\geq 2,k_x}^{\text{warp}, b,t}=\left\langle\psi_{n\geq2,k_x}^{b,t} \right| \mathcal {H}_{2,k_x,y}^{\text{warp},b,t} \left| \psi_{n\geq2,k_x}^{b,t} \right\rangle =0,
\end{equation}
where $| \psi_{n\geq 2,k_x}^{b,t} \rangle$ is given in Eq.~(\ref{eig_vecn}). The first-order energy corrections to the non-degenerate first pseudo Landau levels [i.e., those whose unperturbed dispersions read  $\mathcal E_{1,k_x}^{\text{warp}, b,t}=\pm(\epsilon_{b,t}^2+\gamma_1^2)^{1/2}$] are
\begin{equation} \label{dE_warp1}
 \Delta\mathcal E_{1,k_x}^{\text{warp}, b,t}=\left\langle\psi_{1,k_x}^{b,t} \right| \mathcal {H}_{2,k_x,y}^{\text{warp},b,t} \left| \psi_{1,k_x}^{b,t} \right\rangle =0,
\end{equation}
where $| \psi_{1,k_x}^{b,t} \rangle$ is given in Eq.~(\ref{eig_vec1}) with $c_{1,k_x,A_u}\neq 0$. As the remaining first pseudo Landau level is a flat band (i.e., $\mathcal E_{1,k_x}^{b,t}=0$), degenerate with the zeroth pseudo Landau level, the first-order energy corrections to these two bands must be determined by diagonalizing the matrix
\begin{equation}
\begin{split}
&\Lambda^{\text{warp}}_{2,k_x}=
\\
&
\begingroup
\renewcommand*{\arraystretch}{2}
\begin{bmatrix}
\left\langle\psi_{0,k_x}^{b,t} \right| \mathcal {H}_{2,k_x,y}^{\text{warp},b,t} \left| \psi_{0,k_x}^{b,t} \right\rangle & \left\langle\psi_{0,k_x}^{b,t} \right| \mathcal {H}_{2,k_x,y}^{\text{warp},b,t} \left| \psi_{1,k_x}^{b,t} \right\rangle
\\
\left\langle\psi_{1,k_x}^{b,t} \right| \mathcal {H}_{2,k_x,y}^{\text{warp},b,t} \left| \psi_{0,k_x}^{b,t} \right\rangle & \left\langle\psi_{1,k_x}^{b,t} \right| \mathcal {H}_{2,k_x,y}^{\text{warp},b,t} \left| \psi_{1,k_x}^{b,t} \right\rangle
\end{bmatrix},
\endgroup
\end{split}
\end{equation}
where $| \psi_{1,k_x}^{b,t} \rangle$ is given in Eq.~(\ref{eig_vec1}) but with $c_{1,k_x,A_u}=0$, while $| \psi_{0,k_x}^{b,t} \rangle$ is given in Eq.~(\ref{eig_vec0}). It is straightforward to check that $\Lambda^{\text{warp}}_{2,k_x}$ is a null matrix such that 
\begin{equation} \label{dE_warp10}
\Delta\mathcal E_{1(0),k_x}^{\text{warp}, b,t}=0.
\end{equation} 
As the first-order energy corrections vanish for all the pseudo Landau levels, Eqs.~(\ref{PLLn}),~(\ref{PLL1}), and~(\ref{PLL0}) are still valid dispersions in the presence of second-nearest-neighbor interlayer hoppings between non-dimer sites. We emphasize that such a conclusion requires no knowledge on the explicit forms of the coefficients $c_{n,k_x}$'s.

We then study the strain-induced pseudo Landau levels in the presence of second-nearest-neighbor interlayer hoppings between a dimer site and a non-dimer site. As we have made the approximation $\gamma_4^{b,t}(y)\approx \gamma_4$, this interlayer coupling in the presence of strain is still characterized by Eq.~(\ref{H2coup}). Perform Fourier transform along the $x$ direction, Eq.~(\ref{H2coup}) becomes
\begin{align} \label{H2coup_rib}
H_2^{\text{coup}}=&\sum_{k_x,y}a_{l,k_x,y}^\dagger \left[ 2\gamma_4\cos\left(\frac 1 2 k_x\ell_x\right)+\gamma_4\hat s_{\ell_y}\right] b_{u,k_x,y-\frac{\ell_y}{3}} \nonumber
\\
+&\sum_{k_x,y}a_{u,k_x,y}^\dagger \left[ 2\gamma_4\cos\left(\frac 1 2 k_x\ell_x\right)+\gamma_4\hat s_{-\ell_y}\right] b_{l,k_x,y+\frac{\ell_y}{3}} \nonumber
\\
+&\text{H.c.}.
\end{align}
In the continuum limit, we write $\hat s_{\pm \ell_y}=1\pm\ell_y\frac{d}{dy}=1\pm\frac{\epsilon_{b,t}}{2\gamma_0}(\hat a_{b,t}-\hat a_{b,t}^\dagger)$ as derived above. The kernel of Eq.~(\ref{H2coup_rib}) then becomes
\begin{equation} \label{H2kxwarp}
\mathcal {H}_{2,k_x,y}^{\text{coup},b,t}=
\begin{bmatrix}
0 & 0 & 0 & \mathcal h_{k_x,+}^{\text{coup},b,t}
\\
0 & 0 & \mathcal h_{k_x,+}^{\text{coup},b,t} & 0
\\
0 & \mathcal h_{k_x,-}^{\text{coup},b,t} & 0 & 0
\\
\mathcal h_{k_x,-}^{\text{coup},b,t} & 0 & 0 & 0
\end{bmatrix},
\end{equation}
where $\mathcal h_{k_x,\pm}^{\text{coup},b,t} = 2\gamma_4\cos(\tfrac{1}{2}k_x\ell_x)+\gamma_4 \pm \tfrac{\gamma_4}{2\gamma_0}\epsilon_{b,t}(\hat a_{b,t}-\hat a_{b,t}^\dagger)$ is defined for transparency. Treating $\mathcal {H}_{2,k_x,y}^{\text{coup},b,t}$ as perturbation, the first-order energy corrections to the non-degenerate pseudo Landau level with $n\geq2$ are
\begin{align}\label{dE_coupn}
 &\Delta\mathcal E_{n\geq 2,k_x}^{\text{coup}, b,t}=\left\langle\psi_{n\geq2,k_x}^{b,t} \right| \mathcal {H}_{2,k_x,y}^{\text{coup},b,t} \left| \psi_{n\geq2,k_x}^{b,t} \right\rangle
 \\
&=\frac{\gamma_4}{\gamma_0}\epsilon_{b,t}\left(\sqrt{n}c_{n,k_x,A_l}c_{n,k_x,B_u}+\sqrt{n-1}c_{n,k_x,B_l}c_{n,k_x,A_u}\right), \nonumber
\end{align}
where the coefficients in $| \psi_{n\geq2,k_x}^{b,t} \rangle$  read $c_{n,k_x,A_l}=-(\mathcal E_{n,k_x}^{b,t})^2\gamma_1/\Xi_{n,k_x}^{b,t}$, $c_{n,k_x,B_l}=\epsilon_{b,t} \gamma_1 \mathcal E_{n,k_x}^{b,t} \sqrt{n-1}/\Xi_{n,k_x}^{b,t}$, $c_{n,k_x,A_u}=[\epsilon_{b,t}^2(n-1)-(\mathcal E_{n,k_x}^{b,t})^2] \mathcal E_{n,k_x}^{b,t}/\Xi_{n,k_x}^{b,t}$, and $c_{n,k_x,B_u}=[(\mathcal E_{n,k_x}^{b,t})^2-\epsilon_{b,t}^2(n-1)] \epsilon_{b,t}\sqrt n/\Xi_{n,k_x}^{b,t}$ with $\Xi_{n,k_x}^{b,t}=\{[(\mathcal E_{n,k_x}^{b,t})^2-\epsilon_{b,t}^2(n-1)]^2[(\mathcal E_{n,k_x}^{b,t})^2+n\epsilon_{b,t}^2]+(\mathcal E_{n,k_x}^{b,t})^4\gamma_1^2+(\mathcal E_{n,k_x}^{b,t})^2\epsilon_{b,t}^2\gamma_1^2(n-1)\}^{1/2}$ being the normalization coefficient and $\mathcal E_{n,k_x}^{b,t}$ being the pseudo Landau level dispersions given in Eq.~(\ref{PLLn}). The first-order corrections to the non-degenerate first  pseudo Landau levels are
\begin{equation} \label{dE_coup1}
\begin{split}
\Delta\mathcal E_{1,k_x}^{\text{coup}, b,t}=\left\langle\psi_{1,k_x}^{b,t} \right| \mathcal {H}_{2,k_x,y}^{\text{coup},b,t} \left| \psi_{1,k_x}^{b,t} \right\rangle =-\frac{\epsilon_{b,t}^2\gamma_1\gamma_4}{2\gamma_0(\epsilon_{b,t}^2+\gamma_1^2)},
\end{split}
\end{equation}
where the coefficients in $| \psi_{1,k_x}^{b,t} \rangle$ explicitly read $c_{1,k_x,A_l}=\gamma_1/\sqrt{2(\epsilon_{b,t}^2+\gamma_1^2)}$, $c_{1,k_x,B_l}=0$, $c_{1,k_x,A_u}=\pm1/\sqrt{2}$, and $c_{1,k_x,B_u}=-\epsilon_{b,t}/\sqrt{2(\epsilon_{b,t}^2+\gamma_1^2)}$. As the remaining first pseudo Landau level is degenerate with the zeroth pseudo Landau level, their first-order energy corrections are given by the eigenvalues of the matrix
\begin{equation}
\begin{split}
&\Lambda^{\text{coup}}_{2,k_x}=
\\
&
\begingroup
\renewcommand*{\arraystretch}{2}
\begin{bmatrix}
\left\langle\psi_{0,k_x}^{b,t} \right| \mathcal {H}_{2,k_x,y}^{\text{coup},b,t} \left| \psi_{0,k_x}^{b,t} \right\rangle & \left\langle\psi_{0,k_x}^{b,t} \right| \mathcal {H}_{2,k_x,y}^{\text{coup},b,t} \left| \psi_{1,k_x}^{b,t} \right\rangle
\\
\left\langle\psi_{1,k_x}^{b,t} \right| \mathcal {H}_{2,k_x,y}^{\text{coup},b,t} \left| \psi_{0,k_x}^{b,t} \right\rangle & \left\langle\psi_{1,k_x}^{b,t} \right| \mathcal {H}_{2,k_x,y}^{\text{coup},b,t} \left| \psi_{1,k_x}^{b,t} \right\rangle
\end{bmatrix},
\endgroup
\end{split}
\end{equation}
where $| \psi_{1,k_x}^{b,t} \rangle = (\epsilon_{b,t}\ket 0, 0, 0, \gamma_1\ket 1)^T/\sqrt{\epsilon_{b,t}^2+\gamma_1^2}$ and $| \psi_{0,k_x}^{b,t} \rangle = (0, 0, 0, \ket 0)^T$ such that the entries explicitly read $\langle\psi_{0,k_x}^{b,t}| \mathcal {H}_{2,k_x,y}^{\text{coup},b,t} | \psi_{0,k_x}^{b,t} \rangle=0$, $\langle\psi_{1,k_x}^{b,t}| \mathcal {H}_{2,k_x,y}^{\text{coup},b,t} | \psi_{1,k_x}^{b,t} \rangle=\epsilon_{b,t}^2\gamma_1\gamma_4/[(\epsilon_{b,t}^2+\gamma_1^2)\gamma_0]$, and $\langle\psi_{0,k_x}^{b,t}| \mathcal {H}_{2,k_x,y}^{\text{coup},b,t} | \psi_{1,k_x}^{b,t} \rangle=\langle\psi_{1,k_x}^{b,t}| \mathcal {H}_{2,k_x,y}^{\text{coup},b,t} | \psi_{0,k_x}^{b,t} \rangle=[2\cos(\tfrac 1 2 k_x\ell_x)+1]\epsilon_{b,t}\gamma_4/(\epsilon_{b,t}^2+\gamma_1^2)^{1/2}$. Consequently, the first-order energy corrections are
\begin{equation} \label{dE_coup10}
\begin{split}
\Delta\mathcal E_{1(0),k_x}^{\text{coup},b,t} = \frac{\gamma_1\gamma_4\epsilon_{b,t}^2}{2\gamma_0(\epsilon_{b,t}^2+\gamma_1^2)} \pm \frac{\gamma_1\gamma_4\epsilon_{b,t}^2}{2\gamma_0(\epsilon_{b,t}^2+\gamma_1^2)} \times
\\
\sqrt{1+\frac{4\gamma_0^2(\epsilon_{b,t}^2+\gamma_1^2)}{\gamma_1^2 \epsilon_{b,t}^2}\left[2\cos\left(\frac{1}{2}k_x\ell_x\right)+1\right]^2}.
\end{split}
\end{equation} 
We note that the zeroth and first pseudo Landau levels are no longer dispersionless.

As the more realistic model of Bernal-stacked bilayer graphene contains both types of second-nearest-neighbor interlayer hoppings, the resulting perturbative corrections to the strain-induced pseudo Landau levels read
\begin{equation} \label{dE}
\Delta\mathcal E_{n,k_x}^{b,t}
=\Delta\mathcal E_{n,k_x}^{\text{warp},b,t} + \Delta\mathcal E_{n,k_x}^{\text{coup},b,t} = \Delta\mathcal E_{n,k_x}^{\text{coup},b,t},
\end{equation}
which is explicitly determined by Eq.~(\ref{dE_coupn}) for non-degenerate pseudo Landau levels with $n\geq 2$, Eq.~(\ref{dE_coup1}) for non-degenerate first pseudo Landau levels, and Eq.~(\ref{dE_coup10}) for the degenerate zeroth and first pseudo Landau levels.
 
To substantiate our claims, we perform numerical simulations on the band structures of bent and twisted ribbons of the realistic Bernal-stacked bilayer graphene. For a weakly bent ribbon with $\lambda_b=0.209W^{-1}<\lambda_b^\text{c}$, we compare the numerical spectra [blue, Figs.~\ref{fig8}(b$_1$-b$_3$)] with the analytic pseudo Landau levels  [red and orange, Figs.~\ref{fig8}(b$_1$-b$_3$)] obtained by perturbation under three different situations: (i) Interlayer warping, where the pseudo Landau level dispersions $\mathcal E_{n,k_x}^b+\Delta\mathcal E_{n,k_x}^{\text{warp},b}$ can be determined from Eqs.~(\ref{PLLn}), ~(\ref{PLL1}), ~(\ref{PLL0}), ~(\ref{dE_warpn}),~(\ref{dE_warp1}), and~(\ref{dE_warp10}); (ii) Interlayer coupling, where the pseudo Landau level dispersions $\mathcal E_{n,k_x}^b+\Delta\mathcal E_{n,k_x}^{\text{coup},b}$ can be determined from Eqs.~(\ref{PLLn}), ~(\ref{PLL1}), ~(\ref{PLL0}), ~(\ref{dE_coupn}),~(\ref{dE_coup1}), and~(\ref{dE_coup10}); and (iii) Both warping and coupling, where the pseudo Landau level dispersions $\mathcal E_{n,k_x}^b+\Delta\mathcal E_{n,k_x}^{\text{warp},b}+\Delta\mathcal E_{n,k_x}^{\text{coup},b}$ can be determined from Eqs.~(\ref{PLLn}), ~(\ref{PLL1}), ~(\ref{PLL0}), ~(\ref{dE_warpn}),~(\ref{dE_warp1}), ~(\ref{dE_warp10}), ~(\ref{dE_coupn}), ~(\ref{dE_coup1}), and~(\ref{dE_coup10}). We find that the analytic pseudo Landau levels derived with the perturbation approach indeed well capture the numerics [Figs.~\ref{fig8}(b$_1$-b$_3$)]. For a strongly bent ribbon with $\lambda_b=0.336W^{-1}>\lambda_b^\text{c}$, we find the analytic pseudo Landau levels can still satisfactorily fit the numerics [Figs.~\ref{fig8}(c$_1$-c$_3$)]. Specifically, the fit for the warping case is better than those of the other two cases. This is because the neglected strain-modulated warping, scaled as $\hat a_b$ and $\hat a_b^\dagger$, yields no perturbative corrections, while the neglected strain-modulated coupling, also scaled as $\hat a_b$ and $\hat a_b^\dagger$, does produce perturbative corrections, which become important for strong strain. For a weakly twisted ribbon with $\lambda_t=0.679W^{-1}<\lambda_t^\text{c}$ and a strongly twisted ribbon with $\lambda_t=0.885W^{-1}>\lambda_t^\text{c}$, similar fitting quality is observed [Figs.~\ref{fig8}(d$_1$-d$_3$) and~\ref{fig8}(e$_1$-e$_3$)].

\section{Magnetism in realistic Bernal-stacked bilayer graphene}
\label{app_c}
In the main text, we have demonstrated that the interaction-induced global antiferromagnetism of the strained Bernal-stacked bilayer graphene arises from the \emph{sublattice-polarized} zeroth and first pseudo Landau levels. In presence of second-nearest-neighbor interlayer hoppings $\gamma_3$ and $\gamma_4$, the wave functions of these two pseudo Landau levels in general acquire perturbative corrections from higher ($n\geq 2$) pseudo Landau levels such that the sublattice polarization breaks down. Therefore, the magnetism in realistic Bernal-stacked bilayer graphene can be complicated.

We here only tend to understand the magnetism in a rather conceptual way. On the one hand, we note that the second-nearest-neighbor hopping $\gamma_3$ between non-dimer sites (i.e., $B_l$ and $B_u$) connects two sublattices of opposite net spins for various types of strained Bernal-stacked bilayer graphene [Figs.~\ref{fig6}(a$_6$-d$_6$) and~\ref{fig6}(a$_8$-d$_8$)], similar to the nearest-neighbor hopping $\gamma_1$. This indicates that interlayer antiferromagnetism may be favored and the global antiferromagnetism may persist. On the other hand, the second-nearest-neighbor hopping $\gamma_4$ between a dimer site and a non-dimer site (i.e., $A_l$ and $B_u$ or $A_u$ and $B_l$) connects two sublattices of identical net spins for various types of strained Bernal-stacked bilayer graphene [Figs.~\ref{fig6}(a$_5$-d$_5$) and~\ref{fig6}(a$_8$-d$_8$) or Figs.~\ref{fig6}(a$_6$-d$_6$) and~\ref{fig6}(a$_7$-d$_7$)], favoring ferromagnetism. The simultaneous existence of ferromagnetic and antiferromagnetic interlayer couplings may suggest geometrical frustration. The resulting magnetism then must be determined by a more sophisticated numerical approach.

\begin{figure}[t]
\includegraphics[width = 8.6cm]{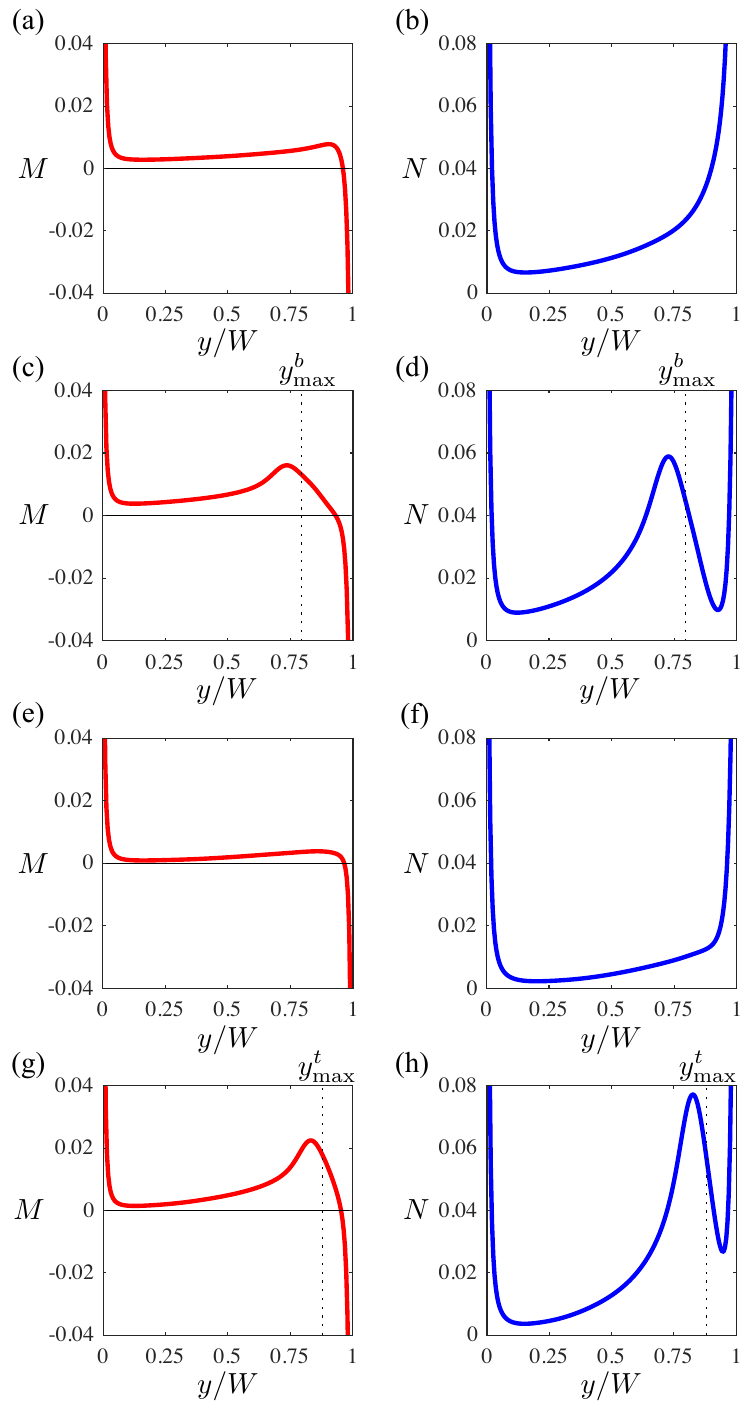}
\caption{Magnetism of a strained ribbon of the realistic Bernal-stacked bilayer graphene with Hubbard interaction. (a, c, e, g) Local ferromagnetic order parameters for various strained ribbons. (b, d, f, h) Local antiferromagnetic order parameters for various strained ribbons. (a, b) Bend $\lambda_b=0.209W^{-1}$. (c, d) Bend $\lambda_b=0.336W^{-1}$. (e, f) Twist $\lambda_t=0.679W^{-1}$. (g, h) Twist $\lambda_t=0.885W^{-1}$. The parameters used are $\gamma_1=0.131|\gamma_0|$ and $\gamma_3=0.115|\gamma_0|$ taken from Ref.~\cite{min2007}. And we set $U=1.2|\gamma_0|$ as in Fig.~\ref{fig7}.
} \label{fig9}
\end{figure}

Nevertheless, we here present a mean-field simulation of the magnetism of the strained Bernal-stacked bilayer graphene with $\gamma_3\neq 0$ but $\gamma_4=0$ (Fig.~\ref{fig9}) such that the interlayer antiferromagnetism holds. We find that both the local ferromagnetic and antiferromagnetic order parameters are similar to those with $\gamma_{3,4}=0$ (Fig.~\ref{fig7}), potentially implying similar global antiferromagnetism. We also find that no reasonable magnetism can be extracted for $\gamma_4\gtrsim0.03|\gamma_0|$, implying the breakdown of the mean-field approach.

\clearpage
\bibliographystyle{apsrev4-1-etal-title_10authors}
\bibliography{strained_BG}
\end{document}